\documentclass[%
reprint,
superscriptaddress,
nofootinbib,
 amsmath,amssymb,
 aps,
prd,
floatfix,
]{revtex4-2}

\usepackage{graphicx}
\usepackage{dcolumn}
\usepackage{bm}
\usepackage[colorlinks,linkcolor=blue]{hyperref}
\usepackage{physics}
\usepackage{booktabs}
\usepackage{tabularx}
\usepackage{tcolorbox}
\usepackage{amsthm}
\usepackage{multirow}
\usepackage{footnote}
\usepackage{soul}
\makesavenoteenv{table}
\usepackage{subcaption}
\usepackage{ragged2e} 
\DeclareCaptionJustification{justified}{\justifying}

\usepackage{tikz}
\usetikzlibrary{arrows,shapes,positioning}
\usetikzlibrary{decorations.markings}
\usetikzlibrary{decorations.pathreplacing,angles,quotes}
\tikzstyle arrowstyle=[scale=1]
\tikzstyle directed=[postaction={decorate,decoration={markings,
    mark=at position 1.0 with {\arrow[arrowstyle]{stealth}}}}]
\tikzstyle reverse directed=[postaction={decorate,decoration={markings,
    mark=at position .0 with {\arrowreversed[arrowstyle]{stealth};}}}]
    
\usepackage{aas_macros}
\newcommand{\caltech}[0]{
    California Institute of Technology, Pasadena, CA 91125, USA
}
\newcommand{\jpl}[0]{
    Jet Propulsion Laboratory, California Institute of Technology, Pasadena, CA 91109, USA
}

\newif\ifshowcomments

\definecolor{brickred}{rgb}{0.8, 0.25, 0.33}

\showcommentstrue

\newcommand{\beq}{\begin{equation}}
\newcommand{\eeq}{\end{equation}}

\newcommand{\bea}{\begin{eqnarray}}
\newcommand{\eea}{\end{eqnarray}}

\begin{document}

\preprint{APS/123-QED}

\title{Constraining effective neutrino species with bispectrum of large scale structures}

\author{Yanlong Shi}
    \email{yanlong@caltech.edu}
    \affiliation{\caltech}
    
\author{Chen Heinrich}%
    \affiliation{\caltech}
\author{Olivier Dor\'e}
    \affiliation{\caltech}
    \affiliation{\jpl}

\date{\today}

\begin{abstract}

Relativistic and free-streaming particles like neutrinos leave imprints in large scale structures (LSS), providing probes of the effective number of neutrino species $N_{\rm eff}$. In this paper, we use the Fisher formalism to forecast $N_{\rm eff}$ constraints from the bispectrum (B) of LSS for current and future galaxy redshift surveys, specifically using information from the baryon acoustic oscillations (BAOs). Modeling the  galaxy bispectrum at the tree-level, we find that adding the bispectrum constraints to current CMB constraints from Planck can improve upon the Planck-only constraints on $N_{\rm eff}$ by about 10\% -- 40\% depending on the survey. Compared to the Planck + power spectrum (P) constraints previously explored in the literature, using Planck+P+B provides a further improvement of about 5\% -- 30\%. Besides using BAO wiggles alone, we also explore using the total information which includes both the wiggles and the broadband information (which is subject to systematics challenges), generally yielding better results. Finally, we exploit the interference feature of the BAOs in the bispectrum to select a subset of triangles with the most information on $N_{\rm eff}$. This allows for the reduction of computational cost while keeping most of the information, as well as for circumventing some of the shortcomings of applying directly to the bispectrum the current wiggle extraction algorithm valid for the power spectrum. In sum, our study validates that the current Planck constraint on $N_{\rm eff}$ can be significantly improved with the aid of galaxy surveys before the next-generation CMB experiments like CMB-Stage 4.
\end{abstract}

\maketitle


\section{Introduction}

Large scale structure (LSS) surveys have proved useful in furthering our understanding of the Universe, constraining various cosmological parameters such as the initial conditions, energy content and evolution of the Universe. Current and future specstrocopic surveys such as BOSS~\cite{DawsonSchlegel2013}, eBOSS~\cite{DawsonKneib2016},
DESI~\cite{DESICollaborationAghamousa2016a,DESICollaborationAghamousa2016b}, Euclid~\cite{LaureijsAmiaux2011}, PFS~\cite{TakadaEllis2014}, SPHEREx~\cite{DoreBock2014}, and Roman~\cite{SpergelGehrels2015} are designed to measure the distribution of galaxies in redshift space, which is especially well-suited for measuring properties of the baryon acoustic oscillations (BAOs)~\cite{EisensteinHu1998,SeoEisenstein2003,BlakeGlazebrook2003,ColePercival2005,EisensteinZehavi2005}.

The BAOs are imprints left behind by the propagation of sound waves inside the  photon-baryon plasma before recombination. They induce a strong correlation of galaxies separated by the sound horizon scale $r_{\rm s}$ ($\sim 100\, h^{-1}{\rm Mpc}$), while in the Fourier space, they show up as oscillatory features with a frequency of roughly $2\pi/r_s$. This sound horizon (also called BAO scale) is sensitive to various cosmological parameters like the baryon and dark matter densities, and can be used to constrain those parameters. Moreover, in the precision cosmology era, it is possible to not only extract information from the frequency of the baryon acoustic oscillations in Fourier space, but also from their amplitude envelope and phases. Studies have shown that phases in the BAO of LSS were able to survive nonlinear and local gravitational evolution~\cite{BaumannGreen2017}, which makes it a novel probe of physical phenomena that alter the BAO phases. 

One notable example of physics that induces phase shifts in the BAO is the effective number of neutrino species $N_{\rm eff}$. This parameter parameterizes the effect on the energy density from any dark radiation relic after the Big Bang, but whose fiducial value in the standard cosmology with three neutrino species is predicted to be 3.046. For various beyond standard model physics such as axions~\cite{PecceiQuinn1977}, light sterile neutrinos~\cite{AbazajianAcero2012} or dark photons~\cite{Holdom1986}, the predictions for $N_{\rm eff}$ would be different, so precision measurements of $N_{\rm eff}$ can provide evidence for either the Standard Model or new physics.

Observationally, a positive deviation from the fiducial value $\Delta N_{\rm eff}$ (either from neutrinos or other light particles) would result in a stronger damping envelope for the BAO wiggles in Fourier space due to diffusion damping~\cite{HouKeisler2013}, which arises from the photon diffusion that erases the anisotropies at scales smaller than the mean free path of photons. Due to neutrinos' free-streaming, it would also affect the dark matter clustering driving the oscillations in the plasma at early times, resulting in a predictable phase shift~\cite{BashinskySeljak2004}. Past studies have used these effects to constrain $N_{\rm eff}$ using either the cosmic microwave background (CMB)~\cite{HouKeisler2013,FollinKnox2015}  or the galaxy power spectrum measurements~\cite{BaumannGreen2018,BaumannBeutler2019}. %

Besides the power spectrum, the bispectrum,  which is the three-point correlation function of the density field in Fourier space, often contains additional information on the cosmological parameters (e.g.,~\cite{YankelevichPorciani2019,IvanovPhilcox2022}).
As the bispectrum describes how densities at three different scales are 
correlated, it is known to be a probe of the primordial density field such as the primordial non-Gaussianity, as well as the late-time nonlinear growth of structures~\cite{DalalDore2008, Scoccimarro2000,SefusattiCrocce2006}.
The BAO scale information is also contained in the bispectrum measurement of spectrosopic surveys and have been detected in the BOSS data using the total bispectrum which includes both the broadband and wiggle components~\cite{PearsonSamushia2018} (as well as in the real-space three-point correlations~\cite{GaztanagaCabre2009,SlepianEisenstein2017a,SlepianEisenstein2017b}). 

Later in Ref.~\cite{ChildTakada2018}, the authors showed that one can also extract the BAO wiggles from the bispectrum instead of using the entire broadband plus wiggles information, using the technique of ``bispectrum interference". The bispectrum interference consists of the interplay of BAO wiggles between different wavevectors in the bispectrum signal, leading to constructive or destructive interferences which are made explicitly manifest in a new parametrization of the triangle configurations.
In this new coordinates, it becomes clear that the BAO information is rather concentrated to some subsets of triangle configurations, which can be used to reduce computational cost. 
More importantly, the interference is sensitive to amplitude and phase shift effects, which makes it ideal for constraining $N_{\rm eff}$.
In this paper, we use the bispectrum interference technique, and apply it for the first time to study the constraints on $N_{\rm eff}$ using the BAO wiggles in the bispectrum. We also investigate, for comparison, the constraints from using the total bispectrum (broadband + wiggles). In both cases, the bispectrum yields better constraints than the power spectrum. Although as in the case of the power spectrum study in Ref.~\cite{BaumannGreen2018}, the current LSS bipectrum constraints themselves are not as competitive as the current CMB constraints, we find that when combined, the Planck + LSS results improve significantly upon the Planck-alone constraints. This can be useful for achieving a better $N_{\rm eff}$ constraints before CMB-Stage 4 (CMB-S4), which would require a more futuristic LSS survey to be improved upon (or possibly modeling to higher $k_{\rm max}$ than our fiducial $k_{\rm max} = 0.2\ h \mathrm{Mpc}^{-1}$ with the upcoming LSS surveys). 
Finally we show that the bispectrum interference is helpful in reducing computational costs by effectively reducing the triangle configurations used.

The paper is structured as follows. In Sec.~\ref{sec:background}, we introduce the background on neutrino physics and their effects on the matter power spectrum and the matter bispectrum; we also review the technique of bispectrum interference. In Sec.~\ref{sec:modeling}, we describe the modeling of our observables, namely the galaxy power spectrum and the galaxy bispectrum. We present the Fisher matrix formalism in Sec. \ref{sec:fisher} used to obtain the forecast constraints. In Sec.~\ref{sec:results} we present the results, comparing the bispectrum to the power spectrum constraints, as well as showing how one can use the bispectrum interference to decrease the computational cost. Finally, in Sec.~\ref{sec:conclusions} we summarize and discuss our conclusions.

Throughout the paper, we use the fiducial $\Lambda$CDM cosmology based on Planck 2018 results with the data \emph{TT,TE,EE+lowE+lensing}~\cite{PlanckCollaborationAghanim2020} with the initial spectrum amplitude and tilt $A_{\rm s}=2.207\times 10^{-9}$ and $n_{\rm s}=0.9645$, the baryon and cold dark matter densities $\omega_{\rm b}\equiv \Omega_{\rm b} h^2=0.0223$ and $\omega_{\rm c}\equiv \Omega_{\rm c} h^2=0.1188$, the sound horizon angular extent at recombination $\theta_\star \equiv r_{\rm s}(z_\star)/D(z_\star) = 1.0411\times 10^{-2}$, and the reionization optical depth $\tau=0.0544$. The resulting fiducial value of the sound horizon at recombination is $r_{\rm s} = 147.49\,{\rm Mpc}$.
Finally the fiducial value of $N_{\rm eff}$ used is 3.046. We use a helium fraction $Y_{\rm p} = 0.239$ that is consistent with BBN results. 

\section{Background}
\label{sec:background}

In this section, we first briefly review the neutrino-induced effects in the matter power spectrum,
before introducing the matter bispectrum and its corresponding response to $N_{\rm eff}$. Then we review the technique of the bispectrum interference developed in Ref.~\cite{ChildTakada2018} and and apply it to the specific case of $N_{\rm eff}$.

\subsection{Effects of $N_{\rm eff}$ on the matter power spectrum}

We now briefly review the neutrino-induced phase shifts in the BAOs and describe the how we model these effects in the matter power spectrum. For more details, we refer the readers to Ref.~\cite{BaumannGreen2018}. 

%

At very early times ({$\sim 1~{\rm s}$ after the Big Bang}), when the temperature of the Universe was high ($\gtrsim 3~{\rm MeV}$), neutrinos were kept in equilibrium with the rest of the plasma; they decoupled from the plasma when their interaction rate decreased below the expansion rate of the Universe. Around the same time, the annihilation of electrons and positrons, and the entropy of these particles was mostly transferred to photons. While this event increased the photon temperature, it did not affect that of the neutrinos as much. Assuming that neutrinos decoupled instantaneously, the neutrino's temperature would have been lower by a factor of $T_{\nu}/T_{\gamma} = (4/11)^{1/3}$ relative to that of the photons. {The effective number of neutrino species $N_{\rm eff}$ is then defined from}
\beq
    \epsilon_\nu = \frac{\rho_\nu}{\rho_{\rm \gamma} + \rho_{\nu}} = \frac{N_{\rm eff}}{\alpha_\nu + N_{\rm eff}},
    \label{equ:epsilon_nu}
\eeq
which is the neutrino energy density relative to the total radiation, and
\beq
    \alpha_{\nu} = 
    \frac{8}{7} \left(\frac{11}{4}\right)^{4/3}.
\eeq

In reality, the neutrino decoupling was not instantaneous; taking this into account along with various QED corrections, we have that $N_{\rm eff} = 3.046$ (corresponding to $\epsilon_{\nu} = 0.405$) in the Standard Model~\cite{Steigman2001, ManganoMiele2005}. Because the presence of any additional light particles that were relativistic at early times would simply add to the effective number of neutrinos measured, detecting deviations from $N_{\rm eff}=3.046$ could be hints of new physics beyond the Standard Model.

Because the free-streaming particles like neutrinos alter the BAO signatures observed in the CMB and in galaxy surveys, BAOs can be used to probe $N_{\rm eff}$. 
These oscillations originate from before recombination, when the photons and baryons were tightly coupled in a photo-baryon plasma (due to the Thomson scattering between photons and free electrons and the Coulomb interactions between electrons and protons). Acoustic oscillations perturbations propagated inside this plasma at the sound speed of $c_{\rm s}\sim c/\sqrt{3}$. When the Universe cooled enough to form stable neutral hydrogen from protons and electrons (at around $T\sim 0.3\, \mathrm{eV}$, $z\sim 1100$), the photons and baryons became decoupled, and the acoustic oscillations froze. This pattern of overdensities frozen in space gave rise to the anisotropies observed in the CMB; they also seeded dark matter perturbations by attracting dark matter, which later caused a preferential formation of galaxies around the sound horizon scale, which is observable in galaxy surveys~\cite{EisensteinZehavi2005,ColePercival2005}.

Since neutrinos had already decoupled from the photon-baryon plasma, they free-streamed at nearly the speed of light which is faster than the sound speed of the plasma at the time of recombination. As a result, their perturbations traveled ahead of the sound waves, altering the gravitational potential perturbations, which is the driving force of the acoustic oscillations~\cite{Baumann2018}. This change left observational signatures that are reflected in both the amplitude and the phases of the acoustic oscillations. The most remarkable effect is a nearly-constant phase shift on small scales proportional to the neutrino energy fraction $\epsilon_{\nu}$~\cite{BashinskySeljak2004,FollinKnox2015,BaumannGreen2018}. %

More specifically, let the comoving matter density contrast be defined as 
$\delta(\vec{x}) = (\rho(\vec{x}) - \bar{\rho})/\bar{\rho}$
where $\bar{\rho}$ is the mean matter density in the Universe.
The matter power spectrum $P_{\rm m}(k)$ is defined as the correlation of the density contrast $\delta(\Vec{k})$ in Fourier space:
\begin{align}
\langle\delta (\Vec{k}) \delta(\Vec{k}') \rangle =(2 \pi)^{3} \delta_{\mathrm{D}}(\Vec{k}+\Vec{k}') P_{\rm  m}\left(k\right),
\end{align}
where the Dirac delta $\delta_{\mathrm{D}}$ arises due to statistical homogeneity and isotropy.

We can decompose the linear matter power spectrum into a smooth (non-wiggle) part $P_{\rm m }^{\rm nw}(k)$ and a wiggle part $P_{\rm m}^{\rm w}(k)$ which contains the BAO, and further define $O(k)$ as the ratio $P_{\rm m}^{\rm w}(k)/P_{\rm m}^{\rm nw}(k)$ such that 
\begin{align}
P_{\rm m}(k)=P_{\rm m }^{\rm nw}(k)\left[1+O\left(k\right)\right].
\label{eq:linear_power_spectrum}
\end{align}

To understand the effects of $N_{\rm eff}$ on the matter power spectrum, let us approximate the oscillatory part as $O(k)=A(k) \sin (r_{\rm s} k + \phi(k))$~\cite{BaumannGreen2018,BaumannBeutler2019}, where $A(k)$ is the scale-dependent amplitude, $r_{\rm s}$ is the sound horizon, and $\phi(k)$ is the phase shift term. 

The most visible impact of $N_{\rm eff}$ is on the damping envelope of the oscillations $A(k)$ as a result of diffusion damping during recombination. The finite mean free path of the Thompson scattering between electrons and photons allows the photons to diffuse and erase anisotropies on that scale. More specifically, the damping effect can be described as an exponential term $\exp(-k^2/k_{\rm d}^2)$ applied to the undamped wiggles in the power spectrum, where $k_{\rm d}$ is the damping scale which is related to the number density $n_{\rm e}$ of the free electrons responsible for scattering the photons. The strength of damping can be characterized with the ratio $r_{\rm d}/r_{\rm s} \propto \sqrt{H/n_{\rm e}}$, where $r_{\rm d}\equiv 2\pi/k_{\rm d}$.
When $a_{\rm eq}$ is fixed, we have $r_{\rm d}/r_{\rm s} \propto \sqrt{1/[n_{\rm e}(1-\epsilon_\nu)]}$~\cite{HouKeisler2013}, which means that the diffusion damping effect is stronger when $N_{\rm eff}$ increases~\cite{BaumannGreen2018}. Moreover, since $n_{\rm e} \propto 1-Y_{\rm p} $, there is a degeneracy between $N_{\rm eff}$ and $Y_{\rm p}$ when contrained from the diffusion damping effects alone; we expect therefore that $N_{\rm eff}$ constraints from BAO wiggles be degraded when $Y_{\rm p}$ is marginalized over~\cite{HouKeisler2013, BaumannGreen2018}. %

Besides the damping envelope, another important effect of $N_{\rm eff}$ is on the  scale-dependent phase shift $\phi(k)$.
{As shown in Ref.~\cite{BaumannGreen2017}, even though there is the nonlinear evolution of the matter density field, the phase shift of BAO in the power spectrum is still a robust probe of additional species of light particles.} If the phase shift effect is only due to $N_{\rm eff}$, it could be used to relieve part of the degeneracy between $N_{\rm eff}$ and $Y_{\rm p}$ mentioned above. %

The authors of Ref.~\cite{BaumannGreen2018} found that the oscillations can be well described as 
\begin{align}
    O^{\rm temp}(k) = O^{\rm fid}\left(\frac{k}{\alpha}+(\beta-1)\frac{f(k)}{r_{\rm s}^{\rm fid}}\right), \label{equ:phase_shift}
\end{align}
where $O^{\rm fid}(k)$ is the oscillatory piece of the power spectrum in the fiducial cosmology. Here $\alpha = r_{\rm s}^{\rm fid}/r_{\rm s}$ accounts for the `stretching"  or ``compressing" the BAO oscillations in Fourier space as the sound horizon may be different in the given cosmology than that of the fiducial cosmology. The additional phase shift due to the deviation from the fiducial $N_{\rm eff}^{\rm fid}$ is proportional to $\beta-1$, where $\beta \equiv \epsilon_\nu/\epsilon_\nu^{\rm fid}$ is normalized such that $\beta=1$ for $N_{\rm eff}=N_{\rm eff}^{\rm fid}$. The function $f(k)$ describes the shape of the scale-dependent phase shift and can be approximated using the template derived from simulations in Ref.~\cite{BaumannGreen2018}
\beq
f(k) = \frac{\phi_\infty}{1+(k_\star/k)^\xi},
\eeq
with $\phi_\infty=0.227$, $k_\star=0.0324\,h\,{\rm Mpc}^{-1}$, and $\xi=0.872$. Later in  Ref~\cite{BaumannBeutler2019}, the amplitude of the phase shift $\beta$ was successfully measured in the BOSS DR12 data {(e.g., $\beta=2.22\pm 0.75$ when marginalizing over the $\Lambda$CDM+$N_{\rm eff}$, using a  prior  on $\alpha$ from Planck)}.

%
%
%
%
%

%
%
%
%

%

%
%
%

%

\begin{figure*}
    \centering
    \includegraphics[width=\linewidth]{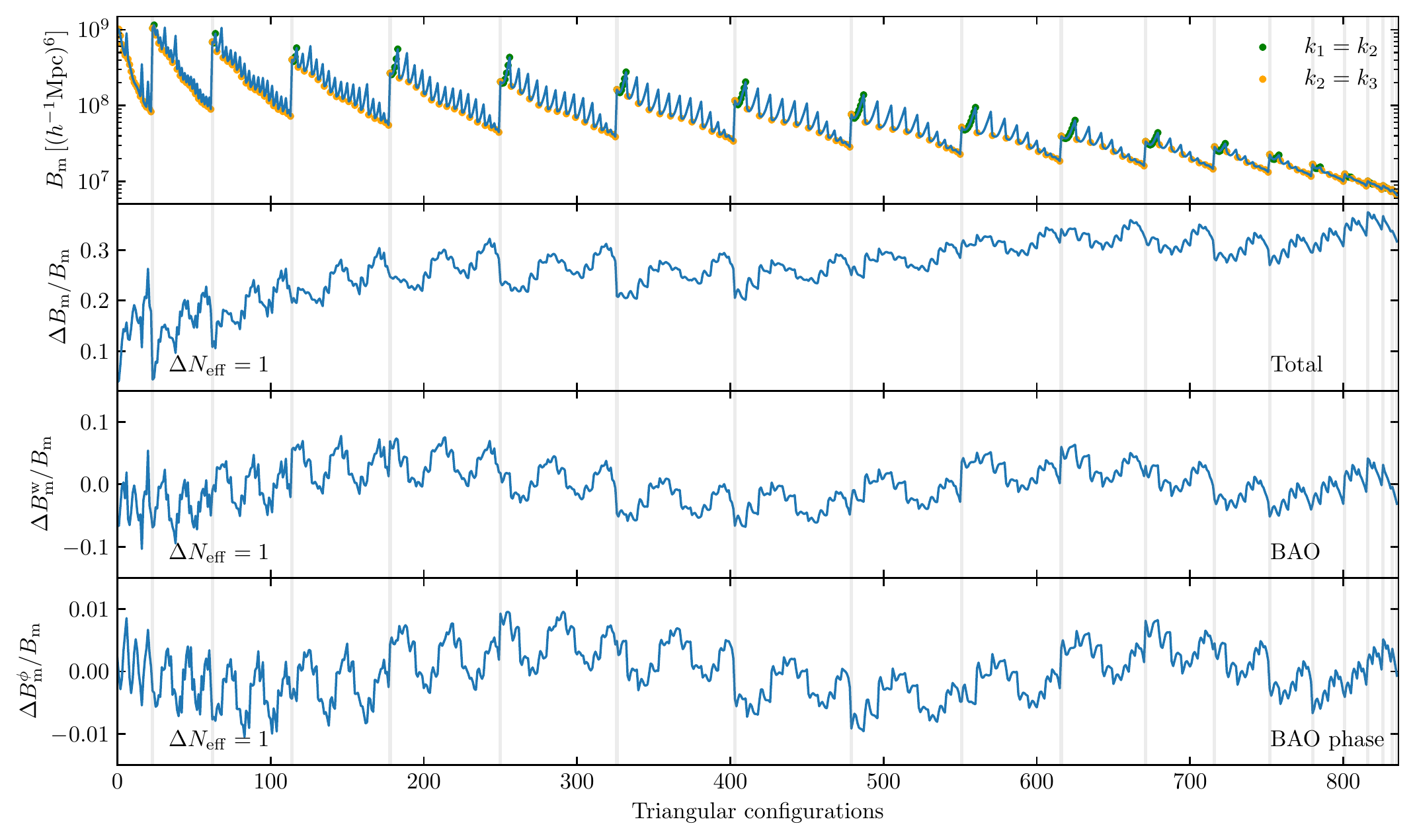}
    \caption{
    The matter bispectrum and the fractional changes of its various components, due to a step $\Delta N_{\rm eff} = 1$ from the fiducial value (with $a_{\rm eq}$ fixed for all the variations in the lower three panels). 
    \emph{Top}: Matter bispectrum at redshift $z=0$.
    \emph{Upper middle}: Fractional change in the total bispectrum including broadband and BAO wiggles effects (see Eq.~\ref{equ:matter-bis-tree}). 
    \emph{Lower middle}: Fractional change in the wiggle part of the matter bispectrum (see Eq.~\ref{equ:matter-bis-wiggles}). This includes both the diffusion damping and the phase shift effects. 
    \emph{Bottom}: Fractional change in the phase shift part of the BAO wiggles where the phase shift is modeled using the template derived in Ref.~\cite{BaumannGreen2018} (see Eq.~\ref{equ:matter-bis-phase}).
    The triangle configurations are ordered first by increasing $k_1$ (steps corresponding to gray lines), then $k_2$ (corresponding to orange dots) and then $k_3$.
    Each $k$ is sampled linearly between $0.01\,h\,{\rm Mpc}^{-1}$ and $0.2 \,h\,{\rm Mpc}^{-1}$ with a bin size $\Delta k=0.01\,h\,{\rm Mpc}^{-1}$ (yielding $n_k = 20$). We impose triangle conditions and count only the unique triangles by imposing the order $k_1 \le k_2 \le k_3$. 
    }
    \label{fig:bispectrum}
\end{figure*}

\subsection{Effects of $N_{\rm eff}$ on the matter bispectrum}

The bispectrum is the three-point function of the density contrast
in Fourier space 
\begin{align}
    \langle\delta (\Vec{k}_{1}) \delta(\Vec{k}_{2}) \delta(\Vec{k}_{3}) \rangle& = (2 \pi)^{3} \delta^{\mathrm{D}}(\Vec{k}_{1}+\Vec{k}_{2}+\Vec{k}_{3}) B_{\rm m}\left(k_{1}, k_{2}, k_{3}\right), 
\end{align}
where $\Vec{k}_1, \Vec{k}_2$ and $\Vec{k}_3$ form a closed triangle. In the standard perturbation theory (SPT)~\cite{BernardeauColombi2002}, the tree-level contribution to the matter bispectrum is
\begin{align}
    B_{\mathrm{m}}(\vec k_1, \vec k_2, \vec k_3) = 2 F_{2}(\Vec{k}_1, \Vec{k}_2) P_{\rm m}(k_1) P_{\rm m}(k_2) + 2\; \mathrm{cyc.,}
    \label{equ:matter-bis-tree}
\end{align}
where 
\begin{align}
    F_{2}(\Vec{k}_1, \Vec{k}_2) = \frac{5}{7} + \frac{\hat{k}_i \cdot \hat{k}_j}{2} \left(\frac{k_i}{k_j} + \frac{k_j}{k_i} \right) + \frac{2}{7}(\hat{k}_i\cdot \hat{k}_j)^2
\end{align}
is the second-order density kernel in SPT, and  $P_{\rm m}^{\rm lin}$ is the linear matter power spectrum. The tree-level expression is only valid in the linear regime for $k\lesssim 0.2\,h\, {\rm Mpc}^{-1}$ {at $z=0$ (see tests shown in Ref.~\cite{BaldaufMercolli2015}). At higher redshifts, the linear regime extends to a higher $k$ since there is less nonlinearity, but we choose $k_{\rm max}=0.2\,h\,{\rm Mpc}^{-1}$ in our conservative forecast.}

In Fig.~\ref{fig:bispectrum} top panel, we show the tree-level matter bispectrum at $z=0$, calculated using the linear matter power spectra from \texttt{CAMB}~\cite{LewisChallinor2000}. We order the triangle configurations first by increasing $k_1$, then by increasing $k_2$, and then by $k_3$. To avoid double-counting, we only include triangle configurations that satisfy $k_1\le k_2 \le k_3$. The grey lines correspond to where $k_1$ steps up and the orange dots, where $k_2$ steps up. The value of $k_3$ increases from one orange dot $k_3=k_2$ until $k_{\rm max}$ before coming back down at the next orange dot. The green dots show increasing $k_3$ for fixed $k_1 = k_2$.

In the lower panels, we examine the changes in the different parts of the matter bispectrum corresponding to a step $\Delta N_{\rm eff}=1$ from its fiducial value. In this process, we keep $a_{\rm eq}$ fixed to break the degeneracy between $N_{\rm eff}$ and $\omega_{\rm c}$. 
The second row shows the change in the total matter bispectrum, which includes both the broadband and the BAO wiggles. 
The third row shows the changes in $B_{\rm m}^{\rm w}$, the wiggle part of the bispectrum $B_{\rm m}^{\rm w} = B_{\rm m} - B_{\rm m}^{\rm nw}$ relative to the total bispectrum, where the non-wiggle bispectrum $B_{\rm m}^{\rm nw}$ is defined as in Eq.~\eqref{equ:matter-bis-tree} but using the smooth part of the matter power spectrum $P_{\rm m}^{\rm nw}$, so that
\begin{align}
    B^{\rm w}_{\rm m} & = 2 F_{2}(\Vec{k}_1, \Vec{k}_2) P_{\rm m}^{\rm nw}(k_1;N_{\rm eff}^{\rm fid}) P_{\rm m}^{\rm nw}(k_2;N_{\rm eff}^{\rm fid})
    \nonumber\\
    & \times [O(k_1;N_{\rm eff}) + O(k_2;N_{\rm eff}) + O(k_1;N_{\rm eff})O(k_2;N_{\rm eff})] \nonumber\\
    & + 2\; \mathrm{cyc.\, perm.}\label{equ:matter-bis-wiggles}
\end{align}

Finally, the last row of Fig.~\ref{fig:bispectrum} shows the change in the phase shift part of the matter bispectrum $B^{\phi}_{\rm m}$ relative to the total matter bispectrum, where
\begin{align}
    B^{\phi}_{\rm m} & = 2 F_{2}(\Vec{k}_1, \Vec{k}_2) P_{\rm m}^{\rm nw}(k_1;\beta^{\rm fid}) P_{\rm m}^{\rm nw}(k_2;\beta^{\rm fid})
    \nonumber\\
    & \times [O^{\rm temp}(k_1;\beta) + O^{\rm temp}(k_2;\beta) \nonumber\\
    & + O^{\rm temp}(k_1;\beta)O^{\rm temp}(k_2;\beta)] \nonumber\\
    & + 2\; \mathrm{cyc.\, perm.}\label{equ:matter-bis-phase}
\end{align}
Here $P^{\rm nw}_{\rm m}(k;\beta^{\rm fid})$ is obtained with the fiducial $N_{\rm eff}$, while $O^{\rm temp}(k;\beta)$ is the template defined in Eq.~\eqref{equ:phase_shift} with $O^{\rm fid}$ fixed, so that varying $\beta(N_{\rm eff})$ in $B^{\phi}_{\rm m}$ represents only the phase-shift effects induced by $N_{\rm eff}$ while ignoring other effects in the BAO wiggles as well as the broadband effects. %

Comparing the three signals, we find that the total fractional change $\Delta B_{\rm m}/B_{\rm m}$ is always positive {since $\Omega_{\rm m}$ increases when we increase $N_{\rm eff}$ but keep $a_{\rm eq}$ fixed, which means the amplitude of matter density fluctuations entering the horizon during the matter-dominated era is larger}. The fractional changes in the BAO wiggles $\Delta B_{\rm m}^{\rm w}/B_{\rm m}$ and in the BAO phase shifts $\Delta B_{\rm m}^{\phi}/B_{\rm m}$ are oscillatory. The amplitude of the deviations are also indicators of the information contained: {The change in the total bispectrum $\Delta B_{\rm m}/B_{\rm m}$} contains all the information one can extract from the matter bispectrum, so it has the largest amplitude, {up to a few times that of the fractional changes illustrated in the third row from wiggles alone.} 

The total signal is generally increasing with larger triangle configuration index which corresponds to going to larger $k$'s, {since it is dominated by the effects of $N_{\rm eff}$ on the broadband matter power spectrum,} while the amplitude in the third row for the wiggle parts stay mostly stable over the range of triangle configurations we consider, but should damp out at high enough $k$ (not shown here) as the BAO wiggles become suppressed. Finally, the phase-induced BAO deviation is an order of magnitude smaller in its overall amplitude than the other two cases, so it is expected to give much less stringent constraints on $N_{\rm eff}$. We will study $N_{\rm eff}$ constraints with the phase shift effect alone in the appendix only for the purpose of literature comparison. 

\begin{figure}
    \centering 
    \begin{subfigure}[b]{0.8\linewidth}
      \begin{tikzpicture}
            \coordinate (A) at (-1.5*3/4,2*3/2.5) ;
            \coordinate (B) at (0,0) ;
            \coordinate (Ap) at (-1.5,2);
            \coordinate (C) at (-2.5,0) ;
            \coordinate (D) at (1,0);
        
            \draw[directed] (B) -- (A);
            \node[] at (-0.8, 1)  {$\vec{k}_2$};
            \draw[directed] (C) -- (B);
            \node[] at (-1.35, 0.25)  {$\vec{k}_1$};
            \draw[directed] (A) -- (C);
            \node[] at (-1.6, 1)  {$\vec{k}_3$};
            \draw[dash pattern=on5pt off3pt] (B) -- (D);
        
            \draw (0.25,0.) arc (0:180-58.13:0.25);
            \node[] at (0.5, 0.2)  {$\theta$};
            \node[] at (1.25, 1)  {$k_2=k_1+\pi \delta / r_{\rm s}$};
        \end{tikzpicture}
    \end{subfigure}%
    \\
    \vspace{.5 cm}
    \begin{subfigure}[b]{0.55\linewidth}
      \begin{tikzpicture}
            \coordinate (A) at (1.5*3/2.5,2*3/2.5) ;
            \coordinate (B) at (0,0) ;
            \coordinate (Ap) at (1.5,2);
            \coordinate (C) at (-2.5,0) ;
            \coordinate (D) at (1,0);
        
            \draw[directed] (B) -- (A);
            \node[] at (0.4, 1)  {$\vec{k}_2$};
            \draw[directed] (C) -- (B);
            \node[] at (-1.35, 0.25)  {$\vec{k}_1$};
            \draw[directed] (A) -- (C);
            \node[] at (-1.35, 1)  {$\vec{k}_3$};
            \draw[dash pattern=on5pt off3pt] (B) -- (D);
        
            \draw (0.25,0.) arc (0:53.13:0.25);
            \node[] at (0.5, 0.2)  {$\theta$};
            \node[] at (2.5, 1)  {$k_2=k_1+\pi \delta / r_{\rm s}$};
        \end{tikzpicture}
    \end{subfigure}%
    \caption{
    An illustration of the Child18 coordinates ($k_1$, $\delta$, $\theta$) (Eq.~\ref{equ:child18}) for two triangle configurations. Top and bottom are examples of a triangle with $\theta>\pi/2$ and $\theta<\pi/2$ respectively, where $\theta$ is the angle between $\vec{k}_1$ and $\vec{k}_2$. As pointed out in Ref.~\cite{ChildTakada2018}, this description of the triangle shape which is an alternative to $(k_1, k_2, k_3)$ can be used to determine whether BAO wiggles are interfering constructively ($\delta$ is even) or destructively ($\delta$ is odd) in the bispectrum.
    %
    %
    }
    \label{fig:coordinate}
\end{figure}
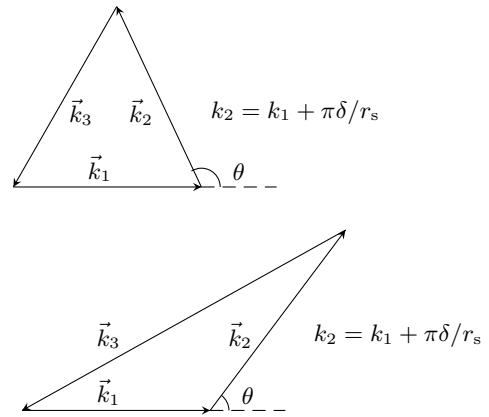

\subsection{Bispectrum interference}
\label{sec:interference}

In this study we will extract information from the BAO wiggles in the bispectrum in order to constrain $N_{\rm eff}$. As we have seen in the previous subsection, the $N_{\rm eff}$ signal in the BAO part of the bispectrum oscillates around zero with
triangle configurations. Using the techniques of bispectrum interference, first introduced in Ref.~\cite{ChildTakada2018}, we will identify later in Section~\ref{sec:notes_on_interference} the subset of triangles that contribute the most to the $N_{\rm eff}$ constraint and show how it can increase computational efficiency. We now briefly review the concept of bispectrum interference and show its effects for the $N_{\rm eff}$. 
In Ref.~\cite{ChildTakada2018}, the authors proposed a new set of coordinates $(k_1, \delta, \theta)$ (hereafter `Child18 coordinate'') where
\begin{align}
    k_2 = k_1+ \pi \delta/r_{\rm s} \qquad \mathrm{and}\qquad \cos\theta = \hat{k}_1 \cdot \hat{k}_2. \label{equ:child18}
\end{align}
The parameter $\delta$ now parametrizes the phase difference between $k_1$ and $k_2$ in terms of the number of half periods given an oscillation frequency of $\pi/r_s$.
The angle $\theta$ is defined as the angle between $\vec k_1$ and $\vec k_2$ 
and is confined to $0 \leq \theta < \pi$. See Fig.~\ref{fig:coordinate} for an example of configurations with $\theta > \pi/2$ and $\theta < \pi/2$. 

{When $k_1, k_2 \ll k_3$, the first of the three permutations}, $2F_2(\vec{k}_1, \vec{k}_2)P_{\rm m}(k_1)P_{\rm m}(k_2)$ dominates over the other cyclic permutations due to the weighting by $F_2(\vec{k}_1, \vec{k}_2)$ (see Fig. 2 of Ref.~\cite{ChildTakada2018}). 
{In this case, omitting the second and third permutations, we can approximate the ratio $O^{\rm bis}$ as }
 \begin{align}
     O^{\rm bis}(k_1, \delta, \theta) \equiv \frac{B^{\rm w}(k_1, \delta, \theta)}{B^{\rm nw}(k_1, \delta, \theta) } \approx O(k_1) + O(k_2) + \mathcal{O}(O^2),
     \label{equ:bispectrum_wiggle}
 \end{align}
where the second-order term in $O$ is negligible since the BAO wiggles are only a small fraction of the broadband matter power spectrum with $O \ll 1$. This prediction can be verified explicitly by plotting $B^{\rm w}/B^{\rm nw}$ in the Child18 coordinates $(k_1, \delta, \theta)$ \cite{ChildTakada2018}.

%
%
%
%
%
%
%

%

%

%

\begin{figure}
    \centering
    \includegraphics[width=\linewidth]{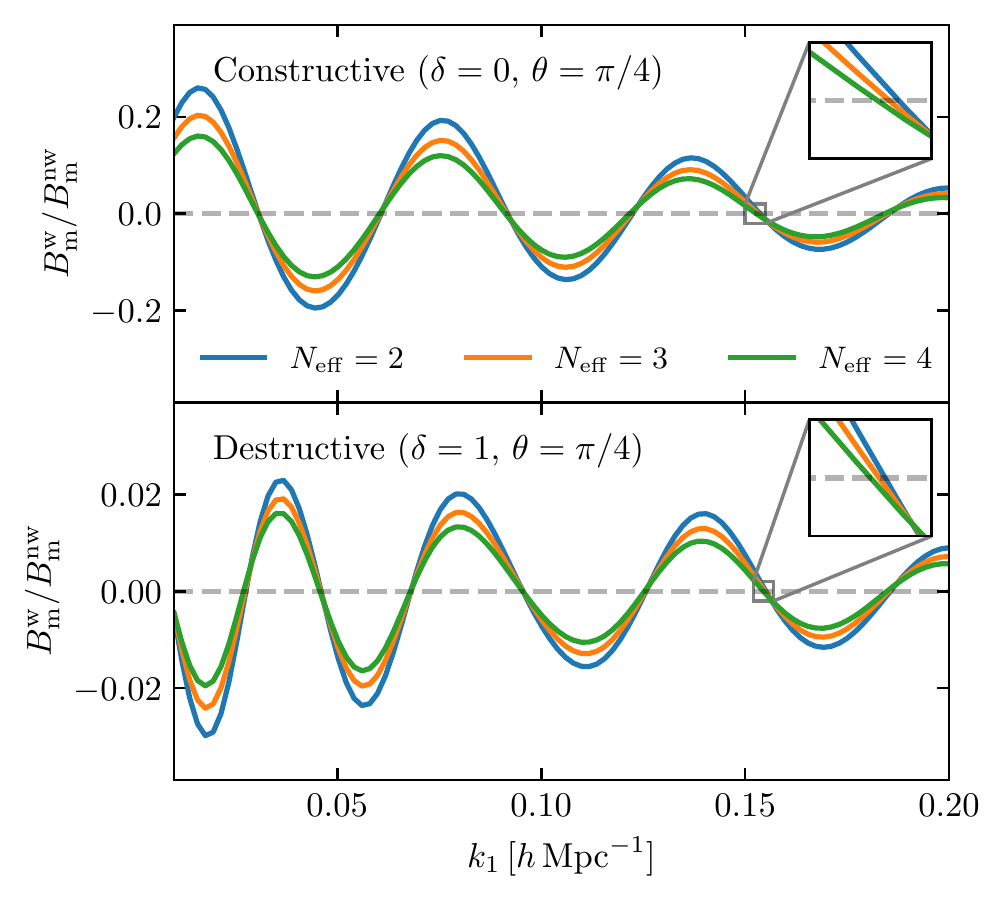}
    \caption{Effects of different $N_{\rm eff}$ on the wiggle part of the matter bispectrum at redshift $z=0$ for triangle configurations with constructive (top, $\delta = 0$) and destructive (bottom, $\delta = 1$) interference. We show $N_{\rm eff} = 2, 3$ and 4 in blue, orange and green lines respectively. 
    The angle $\theta$ between $k_1$ and $k_2$ is fixed at $\theta=\pi/4$. 
    {The matter bispectrum here is computed from the tree-level expression using the linear power spectrum, fixing matter-radiation equality $a_{\rm eq}$ as done in Fig.~\ref{fig:bispectrum}, and further fixing the sound horizon $r_{\rm s}$ to better illustrate the phase shifts}. No damping effects on the wiggles from nonlinear structure formation was included.
    Here the wiggles in the constructive configuration are about an order of magnitude larger than those in the destructive configuration. The two effects from $N_{\rm eff}$ on the wiggles are shown: More diffusion damping of the wiggles with $k$ at larger $N_{\rm eff}$, and higher phase shifts with higher $N_{\rm eff}$. The phase shifts are small enough that our definition of constructive and destructive interference is not affected. 
    }
    \label{fig:beta-bis}
\end{figure}

In Fig.~\ref{fig:beta-bis} we plot $B^{\rm w}/B^{\rm nw}$ as a function of $k_1$ for fixed $\delta$ and $\theta$ to show the effect of bispectrum interference for various values of $N_{\rm eff}$. 
It is clear how the wiggle part of the bispectrum for the constructive triangle configuration ($\delta = 0$) looks significantly different than that of the ``destructive'' configuration ($\delta=1$). 
For our choice of $\theta = \pi/4$, we have $k_1 \leq k_2 \ll k_3$, so the permutations containing $P_{\rm m}(k_3)$ are further suppressed compared to the $(k_1, k_2)$ permutation, and we have that
$B_{\rm m}^{\rm w}/B_{\rm m}^{\rm nw} \approx O(k_1)+O(k_2)$. 
Indeed, we have verified that for the constructive interference in the top panel for which $k_1 = k_2$, the amplitude is approximately twice that of $O(k_1)$. 

In the destructive interference case with $\delta = 1$ in the bottom panel, the cancellation is not perfect since there is a decaying amplitude envelope and the oscillations are not exactly with constant periods, but the amplitude is an order of magnitude lower than that of the constructive interference with the same $\theta$. We show also that shifting $N_{\rm eff}$ by $\pm 1$ around the fiducial value introduces phase shifts that are small enough such that the definition of constructive and destructive interference can still be used largely unaffected using the fiducial $r_s$.

\section{Modeling}
\label{sec:modeling}

We have introduced the effects of $N_{\rm eff}$ on the matter power spectrum and bispectrum, and now we will describe our modeling of the observables, the galaxy power spectrum and bispectrum.

\subsection{Galaxy power spectrum}

We observe the galaxies rather than the matter distribution in the Universe. We model the galaxies as a biased tracer of the underlying matter distribution and account for redshift space distortions (RSD), since we can only measure the galaxy redshifts rather than their true distances. 

We follow Ref.~\cite{BaumannGreen2018} in modeling the observed power spectrum as:
\begin{align}
    P_{\rm g}(\vec{k}) = \frac{Z_1^2(\mu')}{q^3}  P_{\rm m }^{\rm nw}(k') \left(1 + O(k')\mathcal{D}(k', \mu') \right) + \frac{1}{n_{\rm g}},
    \label{eq:galaxy_ps}
\end{align}
where we omitted the redshift bin dependence. Let us now walk through each effect considered. 
\begin{enumerate}
     \item \textit{Redshift space distortions and galaxy bias}

    The linear redshift space distortion effects and the galaxy bias are grouped into one kernel 
    
    \beq
    Z_1(\mu) = b_1 + f\mu^2, 
    \eeq 
    
    where $f(a) =\dd \ln D/\dd \ln a$ is the linear growth rate, and $\mu$ is the cosine of the angle between the line-of-sight vector and the wavevector $\vec{k}$.
    
    We model the galaxy bias to linear order using the linear bias $b_1$ as a function of redshift which we assume to scale as $1/D(z)$, as is appropriate for the evolution of samples in which the galaxy number is conserved:
    \begin{align}
        b_1 (z) = \frac{D(0)}{D(z)} b_1(0). 
        \label{equ:b1}
    \end{align}

    \item \textit{Nonlinear damping of BAO wiggles and its reconstruction}
    
    Baryon acoustic oscillations $O(k)$ are damped by nonlinear structure formation, and we model the damping as
    \begin{eqnarray}
        \mathcal{D}(k, \mu) = \exp\left[-\frac{1}{2}\left(k^2\mu^2\Sigma_{\parallel}^2 + k^2 (1-\mu^2) \Sigma_{\perp}^2 \right) \right]. \notag\\
        \label{equ:bao-damping}
    \end{eqnarray}
    Here $\Sigma_\parallel$ and $\Sigma_\perp$ describe respectively the damping scales for directions parallel and perpendicular to the line-of-sight and they are redshift dependent:
    \begin{align}
        \Sigma_\perp(z) & =  9.4 r\left( \frac{\sigma_8 (z) }{0.9}\right) \, h^{-1}{\rm Mpc}, \notag \\
        \Sigma_\parallel(z) & = \left[1+f(z)\right] \Sigma_\perp(z),
        \label{equ:bao-damping-sigma}
    \end{align}
    where $\sigma_8(z)$ is the variance of the matter density field within 8 $h^{-1}$Mpc at redshift $z$. BAO reconstruction techniques~\cite{EisensteinSeo2007a,PadmanabhanWhite2009,WangYu2017,ShiCautun2018,BirkinLi2019} are often used to revert some of the damping effects due to nonlinear evolution, rendering sharper BAO features.
    Here we model the reconstruction with a fraction $r$: 
    $r = 1$ means no reconstruction, whereas $r = 0$ means full reconstruction. In practice $r$ is modeled as a function of galaxy number density (following Ref.~\cite{BaumannGreen2018} Eq.~3.13) and satisfy $0.5\le r \le 1$. 
    \item \textit{Alcock-Paczynski effect}
    
    A reference cosmology needs to be assumed when converting the observed galaxy redshifts to distances. As a result, in a given cosmology, the different mapping from redshifts to distances results in a different wavenumber $k'$, which is related to $k$ in the fiducial model as
     \begin{align}
        k'(k, \mu, z) & = k\, \sqrt{\frac{\mu^2}{q_{\parallel}^2(z)}+\frac{1-\mu^2}{q_{\perp}^2(z)}}\, ,
        \label{equ:ap} \\
        \mu' (\mu, z) & = \frac{\mu}{q_{\parallel}(z)}/ \sqrt{\frac{\mu^2}{q_{\parallel}^2(z)}+\frac{1-\mu^2}{q_{\perp}^2(z)}}\, . \label{equ:ap_mu}
    \end{align}
    where 
    \bea
    q_{\parallel}(z) &=& D_A(z)/D_A^{\rm ref}(z), \\
    q_{\perp}(z) &=& H^{\rm ref} (z)/H(z).
    \eea
    Here, $D_A$ is the angular diameter distance and $H(z)$ the Hubble parameter.
    All the functions appearing in Eq.~\eqref{eq:galaxy_ps} are evaluated at 
    $k'(k, \mu)$ and $\mu'(\mu)$. 
    Furthermore, a volume factor multiplies the power spectrum due to the different survey volume inferred comoving volumes in the two cosmologies
    \begin{align}
        q = & q_\parallel^{1/3} q_\perp^{2/3}. 
        \label{equ:ap_volume_factor}
    \end{align}
    We choose the reference cosmology to be the same as our fiducial cosmology. We use $h^{-1}\,{\rm Mpc}$ as our distance unit, thus in practice all the AP factors will be adapted by multiplying with $h/h^{\rm ref}$. 

    \item \emph{Systematics}
    
    To account for measurement systematics in the broadband such as stellar contaminations, we add extra terms that are polynomials of $k$ to the non-wiggle power spectrum, and marginalize over their amplitudes~\cite{BaumannGreen2018}
    %
    \begin{align}
        P_{\rm m}^{\rm nw} (k) \to \tilde{B}(k)P_{\rm m}^{\rm nw} + \tilde{A}(k), 
        \label{equ:ps-poly-broadband}
    \end{align}
    where 
    \beq
    \tilde{A}(k) = \sum_{n} \tilde{a}_{n}k^n, \quad \tilde{B}(k) = \sum_{m} \tilde{b}_{m}k^{2m}.
    \label{equ:ps-poly-broadband-coeff}
     \eeq
    In the fiducial case we have $\tilde{b}_0=1$, $\tilde{b}_{m\neq 0}=0$ and $\tilde{a}_n=0$. 
    Note that $\tilde{b}_0$ is degenerate with the linear galaxy bias $b_1$, so we do not vary it in the Fisher forecast.

    For the BAO-only forecast, we use a similar relation on the oscillations to account for systematics such as those arising from the modeling uncertainties in the nonlinear damping of the wiggles (we assumed a particular model in Eq.~\ref{equ:bao-damping} and~\ref{equ:bao-damping-sigma}) or from the wiggle extraction algorithm
    \begin{align}
        O(k) \to B'(k)O(k) + A'(k),\label{equ:ps-poly-wiggle}
    \end{align}
    where $A'(k)$ and $B'(k)$ are defined similarly as in Eq.~\eqref{equ:ps-poly-broadband-coeff}. %
    
    Note that the choice of polynomials terms to marginalize over can sometimes have a significant impact on the result. We study and discuss this more in detail in Sec.~\ref{sec:forecast_dependencies}.
\end{enumerate}

When calculating the covariance of the observed power spectrum, we include the shot noise term $1/n_{\rm g}$ which arises from the sampling of the underlying matter density field with galaxies assuming a Poisson statistics. We assume a constant galaxy density $n_{\rm g}^i$ for the $i$-th redshift bin with middle redshift $z_i$; 
for the case of cosmic variance we set $1/n_{\rm g} = 0$.

\subsection{Galaxy bispectrum}

To model the observed galaxy bispectrum in redshift space, we follow Ref.~~\cite{YankelevichPorciani2019} to include RSD, galaxy biases and the nonlinear damping of BAOs, and use a new set of polynomial terms to account for systematics. The tree-level galaxy bispectrum is modeled as
\begin{align}
    B_{\rm g}(\vec k_1, \vec k_2, \vec k_3) & = 2 \frac{1}{q^6} Z_2(\vec k_1', \vec k_2') Z_1(\vec k_1') Z_1 (\vec k_2')\nonumber\\
    & \times P_{\rm m }^{\rm nw}(k'_1) \left(1 + O(k'_1)\mathcal{D}(k'_1, \mu'_1) \right) \nonumber\\
    & \times P_{\rm m }^{\rm nw}(k'_2) \left(1 + O(k'_2)\mathcal{D}(k'_2, \mu'_2) \right) \nonumber\\ & + \mathrm{2 \ cyc.\ perm.}  \label{equ:galaxy-bispectrum}
\end{align}
{Note that here each $\vec{k}'_i$ ($i=1,2,3$) is corrected from the original $\vec{k}_i$ due to the Alcock-Paczynski effect. We also defined $\mu_i' = \hat{k}'\cdot \hat{s}$ and $k_i'=|\vec{k}_i'|$. The correction follows Eqs.~\eqref{equ:ap} and \eqref{equ:ap_mu}, so $k_i' = k_i'(k_i, \mu_i, z)$ and $\mu_i' = \mu_i'(\mu_i, z)$. More specifically,
\begin{align}
    k_i'(k_i, \mu_i, z) & = k_i\, \sqrt{\frac{\mu_i^2}{q_{\parallel}^2(z)}+\frac{1-\mu_i^2}{q_{\perp}^2(z)}}\, , \\
    \mu_i' (\mu_i, z) & = \frac{\mu_i}{q_{\parallel}(z)}/ \sqrt{\frac{\mu_i^2}{q_{\parallel}^2(z)}+\frac{1-\mu_i^2}{q_{\perp}^2(z)}}\, .
\end{align}
where the $q$ factors are defined the same way as that in the power spectrum for each redshift bin. As also suggested in Eq.~\eqref{equ:galaxy-bispectrum}, in our codes we first convert each $\vec k_i$ into $\vec k_i'$ before 
computing quantities like
$Z_2$, $P_{\rm m}^{\rm nw}$, $\mathcal{O}$, and $\mathcal{D}$ at these wavevectors.
}

The redshift kernel $Z_2$ encodes the RSD and the 
second-order bias effects:
\begin{align}
    Z_2 (\vec k_i, \vec k_j)  = \frac{b_2}{2} + b_1 F_2 (\vec k_i,\vec k_j) + f \mu_{ij}^2 G_2(\vec k_i, \vec k_j) \nonumber \\
     + \frac{f \mu_{ij} k_{ij}}{2} \left[\frac{\mu_i}{k_i} Z_1(k_j) + \frac{\mu_j}{k_j} Z_1(k_i) \right] + \frac{b_{s^2}}{2} S_2(\vec k_i, \vec k_j),
\end{align}
where $\mu_{i} = \vec k_{i} \cdot \hat{n}$, $\vec k_{ij} = \vec k_i + \vec k_j $ and $\mu_{ij} = \vec k_{ij} \cdot \hat{n}$, and where $G_2$ is the second-order kernel of velocity divergence in SPT and $S_2$ is the tidal tensor:
\begin{align}
    G_{2}(\Vec{k}_1, \Vec{k}_2) & = \frac{3}{7} + \frac{\hat{k}_i \cdot \hat{k}_j}{2} \left(\frac{k_i}{k_j} + \frac{k_j}{k_i} \right) + \frac{4}{7}(\hat{k}_i\cdot \hat{k}_j)^2,\\
    S_2(\vec{k}_1, \vec{k}_2) & = (\hat{k}_1\cdot \hat{k}_2)^2 - \frac{1}{3}.
\end{align}

The second-order bias $b_2$ in the fiducial model is calculated using the relation fit from simulations~\cite{LazeyrasWagner2016} 
\begin{align}
    b_2 (z) & = 0.412 - 2.143 \ b_1(z) + 0.929 \ b_1^2 (z)\nonumber\\& + 0.008 \ b_1^3(z),
    \label{equ:b2}
\end{align}
and the fiducial tidal bias $b_{s^2}$ is modeled with~\cite{SaitoBaldauf2014}
\begin{align}
    b_{s^2} (z) = \frac{4}{7}\left(1-b_1(z)\right).
    \label{equ:bs2}
\end{align}
Both are evaluated at the center of the redshift bin and assumed to be constant within the bin.

Note that in addition to the linear galaxy bias in the power spectrum, we also model the second-order bias contributions to the bispectrum in order to account for all contributions at tree-level. This is not consistent with the power spectrum in the sense that we are not cutting the galaxy density at the same order in $\delta_m^{(n)}$ in perturbation theory to model our observables (say by including including one-loop terms in the power spectrum induced by second-order terms in $\delta_g$). But in terms of modeling the lowest-order contributions as a good approximation for each observable in the linear regime, this is a reasonable choice.

Similarly to the power spectrum, the nonlinear damping of BAOs is accounted for by multiplying the wiggle part of  matter power spectrum $O(k)$ by a damping factor $\mathcal{D}$ (Eq.~\ref{equ:bao-damping}). The AP effect is also included just as in the power spectrum: Each wavevector $\vec{k}_i$ is mapped to $\vec{k}_i'$ following Eq. \eqref{equ:ap} and there is a volume factor $1/q^6$ that is different than in the power spectrum, where $q$ was defined in Eq.~\ref{equ:ap_volume_factor}.

To mimic the effects of marginalizing over systematics in the measurements of the broadband bispectrum, we introduce a new set of polynomials $\tilde{A}$ and $\tilde{B}$, different from those used for the power spectrum, such that the galaxy bispectrum becomes %
\begin{align}
B_{\rm g}({\vec k_1, \vec k_2, \vec k_3}) \rightarrow \tilde{B}(k_1, k_2, k_3) B_{\rm g}({\vec k_1, \vec k_2, \vec k_3}) + \tilde{A}(k_1, k_2, k_3).
\end{align}
They allow for different powers of $k_1$, $k_2$ and $k_3$ and are composed of terms proportional to $k_1^r k_2^s k_3^t$: %
\bea
    \tilde{A}(k_1, k_2, k_3) &=& \sum_{n=0} \sum_{(r, s, t) \in S(n)} \tilde{a}_{n}^{rst} \left(k_1^r k_2^s k_3^t + {\rm 2\ cyc.\ perm.}\right), \notag\\
    \tilde{B}(k_1, k_2, k_3) &=& \sum_{n=1} \sum_{(r, s, t) \in S(n)} \tilde{b}_{n}^{rst} \left(k_1^r k_2^s k_3^t + {\rm 2\ cyc.\ perm.}\right). \notag\\
    \label{equ:bs-poly-broadband}
\eea
At each total power $n=r+s+t$, we sum over all $r, s$ and $t$ combinations in the set $S(n) = \{(r, s, t)\, |\, r+s+t =n;\, r\ge s \ge t\}$. For example, $S(3) = \{(3,0,0), (2, 1, 0), (1, 1, 1)\}$. For each $(r,s,t)$ combination, all cyclic permutations of the term are included and assumed to be affected in the same way, being assigned to the same coefficient $\tilde{a}_n^{rst}$. This is purely for simplicity, as there could be systematics terms that affect the different permutations differently. 

As suggested in Ref.~\cite{ChildTakada2018}, one can also extract the BAO wiggles from the bispectrum using the interference coordinates $(k_1, \delta, \theta)$. Recall that $O^{\rm bis}(k_1,\delta, \theta)$ from Eq.~\eqref{equ:bispectrum_wiggle} is the equivalent of the fractional BAO contribution of the BAO wiggles in the bispectrum. It could have additional terms than those shown in the right hand side of Eq.~\eqref{equ:bispectrum_wiggle} for a general triangle shape where the two other cyclic permutations also contribute significantly. We assume that an algorithm can be developed to successfully extract $O^{\rm bis}(k_1,\delta, \theta)$ for all configurations (for a list of necessary problems to solve to reach this assumption, see Appendix~\ref{app:bao-extraction}.) 
%

Working under this assumption that BAO wiggles can be successfully extracted to match the theory $O^{\rm bis}$, we now use a similar technique to marginalize over the modeling uncertainties in the nonlinear damping term for the bispectrum measurement of BAO wiggles, as well as those that possibly arise during the wiggle extraction procedure for the bispectrum. Like in the case for the power spectrum, we apply the polynomials $\tilde{A}'(k_1)$ and $\tilde{B}'(k_1)$ to the undamped part of the oscillations:
\begin{align}
    O^{\rm bis}(k_1,\delta, \theta) \to \mathcal{D}(k_1, \mu_1) \left[\tilde{B}'(k_1) O^{\rm bis}(k_1, \delta,\theta) + \tilde{A}'(k_1) \right], 
\end{align}
where $\mathcal{D}(k_1, \mu_1)$ accounts for the damping of wiggles at $k_1$ due to the nonlinear evolution. Note that this is not an exact treatment, since $O^{\rm bis}(k_1, \delta, \theta) \approx O(k_1) + O(k_1 + \delta)$ to first order in $O$ if the first cyclic permutation in the tree-level expression dominates, so the damping treatment applied as a function of $k_1$ (and but not $k_1 + \delta$) here is only approximate. Yet we expect that marginalizing over the polynomials $\tilde{A}'$ and $\tilde{B}'$ would account for some of the uncertainties in the damping treatment as in the power spectrum case. The definition for $\tilde{A}'(k)$ and $\tilde{B}'(k)$ is just as in the power spectrum case: 
\beq
\tilde{A}'(k)=\sum_n \tilde{a}'_n k^n; \quad \tilde{B}'(k)=\sum_m \tilde{b}'_m k^{2m}.\label{equ:bs-poly-wiggle}
\eeq

In the rest of this paper, we will drop for simplicity the tilde and prime symbols that are used to distinguish between the $a_n$ and $b_m$ coefficients of the various observables, but they should still be distinguishable through the context.

\section{Fisher matrix setups}
\label{sec:fisher}

In this section, we use the Fisher matrix formalism to study the constraining power on $N_{\rm eff}$ and other cosmological parameters from the power spectrum, bispectrum and their combination.

 Let us call the data vector $\vec{d}$ a series of data taken by the observer. The data can be modeled given a set of theory parameters $\vec{p}$ (which we call parameter vector) with a likelihood function $\mathcal{L}(\vec{d}|\vec{p})$. For unbiased estimation of the parameter vectors from the data, denoted $\hat{p}$, the variance $\mathrm{Var}(\hat{p})$ is constrained by the Cramer-Rao inequality $\mathrm{Var}(\hat{p}_i) \ge (-\partial^2 \ln \mathcal{L}(\vec{d}|\vec{p}) / \partial p_i^2 )^{-1}$. In the limit where the likelihood function is well-approximated by a Gaussian, the Fisher matrix
\begin{align}
    F_{ij} = - \frac{\partial^2 \ln \mathcal{L}(\vec{d}|\vec{p})}{\partial p_i \partial p_j} =  \frac{\partial \vec{d}}{\partial p_i} C^{-1} \frac{\partial \vec{d}}{\partial p_j}, \label{equ:fisher-def}
\end{align}
where $C_{ab} = \mathrm{Cov}(d_a, d_b)$ is the covariance matrix of the data vector, gives the best possible constraint on the parameter $i$: $\sigma_{p_i} \ge \sqrt{(F^{-1})_{ii}}$.

\subsection{Power spectrum}

Using the power spectrum as our data vector, and for a single survey volume in which the galaxy number density $\bar{n}_{\rm g}$ is constant, the Fisher matrix is given by~\cite{BaumannGreen2018,YankelevichPorciani2019}
\begin{align}
    F_{i j}=\int_{-1}^{1} \frac{\mathrm{d} \mu}{2} \int_{k_{\min }}^{k_{\max }} \frac{\mathrm{d} k k^{2}}{(2 \pi)^{2}} \frac{\partial \ln P_{\rm g}(k, \mu)}{\partial p_{i}} \frac{\partial \ln P_{\rm g}(k, \mu)}{\partial p_{j}} V_{\mathrm{eff}},
    \label{equ:fisher-power-spectrum}
\end{align}
where
\begin{align}
    V_{\mathrm{eff}} = \left(\frac{\bar{n}_{\rm g} P_{\rm g}(k, \mu)}{\bar{n}_{\rm g} P_{\rm g}(k, \mu)+1}\right)^{2} V,
\end{align}
is the effective survey volume (which is smaller than the real comoving volume $V$), and where we also assumed Gaussian covariance matrix.

In realistic surveys, one often measures the power spectrum in multiple redshift bins. We will treat such cases with the galaxy number density assumed to be constant within each bin, but different from bin to bin. We also assume that there is no correlation between galaxies of different redshift bins, in which case the total Fisher matrix is just the sum over that of all the redshift bins.

To evaluate the Fisher matrix, we need to compute the derivatives of the galaxy power spectrum with respect to the cosmological, bias and systematics parameters. We consider two different ways of evaluating the derivatives and call them loosely here the total
and BAO wiggles, reflecting where the information is drawn from. 
The authors of Ref.~\cite{BaumannGreen2018} also derived constraints on $N_{\rm eff}$ from the phase of the BAO wiggles (see also Ref.~\cite{BaumannBeutler2019}); for the sake of comparison with previous literature, we also include this method in Appendix \ref{app:phase-shift} along with its description and constraints from the power spectrum and bispectrum. %
We now detail the total and BAO wiggle measurements which we focus on in the main text.

\subsubsection{Total constraints}

We start by considering the effect of the parameters $p_i$ on the entire power spectrum including both the broadband shape and the BAO wiggles. 
Our parameter vector is
\begin{align}
    \vec{p}=(N_{\rm eff}, \theta_\star, \omega_b, \omega_c, A_s, n_s, \tau, Y_p;\; \vec b_1;\; {a}_n, {b}_m). \nonumber
\end{align}
The first set of parameters are the cosmological parameters that we label $\Lambda$CDM+$Y_{\rm p}$+$N_{\rm eff}$. We use \texttt{CAMB}~\cite{LewisChallinor2000} to evaluate their derivatives numerically. To do so, we change the fiducial value of each parameter one at a time by a step size of $\pm h$: 
\beq
\frac{\partial P_g}{\partial p_i} \approx \frac{[P_g(p_i=p_i^{\rm fid}+h))-P_g(p_i=p_i^{\rm fid}-h)]}{2h}.
\eeq
Note that we did not fix $\theta_\star$ or $a_{\rm eq}$ here when varying $N_{\rm eff}$.

The second set of parameters are the galaxy biases. We treat them as independent parameters from different redshift bins. For example, if we have $n_z$ redshift bins in a survey, then there are $n_z$ bias parameters in the Fisher matrix. %

Finally, for the polynomial coefficients ${a}_n$ and ${b}_m$ (defined in Eq.~\ref{equ:ps-poly-broadband})
the derivatives are calculated analytically. The polynomial terms are also dependent on the redshift bin, so there will also be $n_z$ coefficients to consider for every $n$ or $m$ in ${a}_n$ or ${b}_m$. For our fiducial setup, we choose $b_{m\le 1}$ following Ref.~\cite{BaumannGreen2018}, amounting to $n_z$ polynomial parameters in the Fisher matrix. Later in Section~\ref{sec:results} we will explore the effects of using a different set of polynomial parameters on our results.

In realistic surveys, the broadband measurement is often prone to systematics like stellar contaminations and those in our modeling of the nonlinear evolution effects and galaxy bias~\cite{DesjacquesJeong2018}. On the other hand, the scale and phase of BAO wiggles are more robust to nonlinear evolution~\cite{EisensteinSeo2007b}, which can also be partly reversed by reconstruction techniques~\cite{EisensteinSeo2007a,PadmanabhanWhite2009,WangYu2017,ShiCautun2018}. For this reason, we will also consider extracting information from only the BAO wiggles in the next subsection.

For our fiducial forecast for the total power spectrum, we set $k_{\rm min}=0.01h\,{\rm Mpc}^{-1}$ and $k_{\rm max} = 0.2h\,{\rm Mpc}^{-1}$.

\subsubsection{BAO wiggles}

We now consider using the information from BAO wiggles alone. Here we take the data vector as the wiggle part of the power spectrum, so that the derivatives are computed as follows, keeping only terms with $P_{\rm g}^{\rm nw}$ fixed at the fiducial cosmology while calculating the derivatives $\partial O(k|\vec{p})/\partial p_i$ numerically:
\begin{align}
    \frac{\partial P^{\rm w}_{\rm g}}{\partial p_i} = P_{\rm g}^{\rm nw} (k, \mu) \mathcal{D}(k, \mu)\frac{\partial O(k|\vec{p})}{\partial p_i}.
    \label{equ:power-spectrum-derivative-bao}
\end{align}
To do so, we need to calculate the matter power spectrum with different cosmological parameters, and separate smooth and oscillatory parts. We follow Ref.~\cite{BaumannGreen2018} to extract the non-wiggle power spectrum by applying a discrete sine transform, cutting the characteristic ``bump'' of the BAO before doing an inverse discrete sine transform. Details of the algorithm may be found in Appendix C of Ref.~\cite{BaumannGreen2018}. 

For power spectrum BAO wiggles, we set $k_{\rm min}=0.01h\,{\rm Mpc}^{-1}$ and  $k_{\rm max} = 0.5h\,{\rm Mpc}^{-1}$. The polynomial terms to include are $a_{n\le 3}, b_{ m\le 4}$~\cite{BaumannGreen2018}.

\begin{table}[!htbp]
\begin{subtable}{\linewidth}
\begin{tabularx}{\textwidth}{XXXXXXX}
\toprule
$z_{\rm mid}$ &0.050 &0.150 &0.250 &0.350 &0.450 &0.550 \\
$10^3n_{\rm g}$ &0.289 &0.290 &0.300 &0.304 &0.276 &0.323 \\
\midrule
$z_{\rm mid}$ &0.650 &0.750 &  &  &  &  \\
$10^3n_{\rm g}$ &0.120 &0.010 &  &  &  &  \\
\bottomrule
\end{tabularx}
\caption{BOSS: $f_{\rm sky}=0.242$ (10000 deg$^2$).}
\end{subtable}

\medskip
 
\begin{subtable}{\linewidth}
\begin{tabularx}{\textwidth}{XXXXXXX}
\toprule
$z_{\rm mid}$ &0.150 &0.250 &0.350 &0.450 &0.550 &0.650 \\
$10^3n_{\rm g}$ &2.380 &1.070 &0.684 &0.568 &0.600 &0.696 \\
\midrule
$z_{\rm mid}$ &0.750 &0.850 &0.950 &1.050 &1.150 &1.250 \\
$10^3n_{\rm g}$ &0.810 &0.720 &0.560 &0.520 &0.510 &0.450 \\
\midrule
$z_{\rm mid}$ &1.350 &1.450 &1.550 &1.650 &1.750 &1.850 \\
$10^3n_{\rm g}$ &0.360 &0.240 &0.130 &0.070 &0.030 &0.010 \\
\bottomrule
\end{tabularx}
\caption{DESI: $f_{\rm sky}=0.339$ (14000 deg$^2$).}
\end{subtable}

\medskip
 
\begin{subtable}{\linewidth}
\begin{tabularx}{\textwidth}{XXXXXXX}
\toprule
$z_{\rm mid}$ &0.650 &0.750 &0.850 &0.950 &1.050 &1.150 \\
$10^3n_{\rm g}$ &0.640 &1.460 &1.630 &1.500 &1.330 &1.140 \\
\midrule
$z_{\rm mid}$ &1.250 &1.350 &1.450 &1.550 &1.650 &1.750 \\
$10^3n_{\rm g}$ &1.000 &0.840 &0.650 &0.510 &0.360 &0.250 \\
\midrule
$z_{\rm mid}$ &1.850 &1.950 &2.050 &  &  &  \\
$10^3n_{\rm g}$ &0.150 &0.090 &0.070 &  &  &  \\
\bottomrule
\end{tabularx}
\caption{Euclid: $f_{\rm sky}=0.364$ (15000 deg$^2$).}
\end{subtable}

\medskip
 
\begin{subtable}{\linewidth}
\begin{tabularx}{\textwidth}{XXXXXXX}
\toprule
$z_{\rm mid}$ &0.700 &0.900 &1.100 &1.300 &1.500 &  \\
$10^3n_{\rm g}$ &0.300 &0.300 &0.400 &0.400 &0.400 &  \\
\bottomrule
\end{tabularx}
\caption{PFS: $f_{\rm sky}=0.048$ (2000 deg$^2$).}
\end{subtable}

\medskip
 
\begin{subtable}{\linewidth}
\begin{tabularx}{\textwidth}{XXXXXXX}
\toprule
$z_{\rm mid}$ &0.100 &0.300 &0.500 &0.700 &0.900 &1.300 \\
$10^3n_{\rm g}$ &9.970 &4.110 &0.501 &0.071 &0.032 &0.016 \\
\midrule
$z_{\rm mid}$ &1.900 &2.500 &3.100 &3.700 &4.300 &  \\
$10^3n_{\rm g}$ &0.004 &0.001 &0.002 &0.002 &0.001 &  \\
\bottomrule
\end{tabularx}
\caption{SPHEREx: $f_{\rm sky}=0.750$ (30940 deg$^2$).}
\end{subtable}

\medskip
 
\begin{subtable}{\linewidth}
\begin{tabularx}{\textwidth}{XXXXXXX}
\toprule
$z_{\rm mid}$ &0.425 &0.475 &0.525 &0.575 &0.625 &0.675 \\
$10^3n_{\rm g}$ &0.482 &0.638 &0.862 &0.975 &1.134 &1.242 \\
\midrule
$z_{\rm mid}$ &0.725 &0.775 &0.825 &0.875 &0.925 &0.975 \\
$10^3n_{\rm g}$ &1.266 &1.282 &1.248 &1.224 &1.189 &1.120 \\
\midrule
$z_{\rm mid}$ &1.025 &1.075 &1.125 &1.175 &1.225 &1.275 \\
$10^3n_{\rm g}$ &1.053 &0.984 &0.903 &0.842 &0.769 &0.713 \\
\midrule
$z_{\rm mid}$ &1.325 &1.375 &1.425 &1.475 &1.525 &1.575 \\
$10^3n_{\rm g}$ &0.645 &0.604 &0.542 &0.487 &0.439 &0.394 \\
\midrule
$z_{\rm mid}$ &1.625 &1.675 &1.725 &1.775 &1.825 &  \\
$10^3n_{\rm g}$ &0.347 &0.309 &0.260 &0.217 &0.187 &  \\
\bottomrule
\end{tabularx}
\caption{Roman: $f_{\rm sky}=0.048$ (2000 deg$^2$).}
\end{subtable}

\caption{Survey specifications used in this study. Here we list the galaxy number density ($n_{\rm g}$, in units of $({\rm Mpc}/h)^{-3}$) as a function of median redshift ($z_{\rm mid}$) at different redshift bins, as well as the sky coverage $f_{\rm sky}$. }
\label{tab:surveys}
\end{table}

\subsection{Bispectrum}

For the bispectrum , the Fisher matrix of a single redshift bin with volume $V$ is given by~\cite{YankelevichPorciani2019,IvanovPhilcox2022}
\begin{align}
    F_{ij} & =  \int_{k_{\rm min}}^{k_{\rm max}} \dd k_1 \int_{k_{1}}^{k_{\rm max}} \dd k_2 \int_{k_{2}}^{k_{\rm max}}\dd k_3 \int_{-1}^{1}\dd \mu_1 \int_{-1}^1 \dd \mu_2\nonumber \\ &  \frac{\partial B }{\partial p_i}\frac{\partial B}{\partial p_j} \frac{V k_1 k_2 k_3 \gamma(\cos \theta)\Sigma (\mu_1, \mu_2, \cos \theta)   }{8 \pi^4 s_{123} P(\vec k_1)P(\vec k_2)P(\vec k_3)}, 
    \label{equ:fisher-bispectrum}
\end{align}
where $k_{\rm min}=0.01h\,{\rm Mpc}^{-1}$ and  $k_{\rm max} = 0.2h\,{\rm Mpc}^{-1}$ in our fiducial setup for both bispectrum broadband and BAO constraints. Note that we use $k_1 \leq k_2 \leq k_3$ in order to count only a unique set of triangles. 
Similar to the power spectrum, we only consider the Gaussian contribution to the covariance matrix (for details, see Appendix \ref{app:N_tri}).  

Here $k_i = |\vec{k}_i|$ and $\mu_i = \hat{k}_i \cdot \hat{n}$ ($i=1,2,3$) where $\hat{n}$ is the line-of-sight direction. The factor $\gamma(\cos\theta)$ describes contributions of different combinations of $(k_1, k_2, k_3)$, and the angular factor $\Sigma (\mu_1, \mu_2, \cos \theta)$ accounts for the orientation of the triangle configuration in the redshift space (see Appendix \ref{app:N_tri} for explicit expressions). Finally, the factor $s_{123}$ accounts for the symmetry factor for different types of triangle configurations: 1, 2 and 6 for the scalene, isosceles and equilateral triangles respectively.

Here again, we have two types of derivatives, one that uses the total information from the broadband and the BAO wiggles, and one that solely extracts information from the wiggles. For the total, we differentiate Eq.~\eqref{equ:galaxy-bispectrum}. For the BAO wiggles, the derivatives $\partial B^{\rm w}_{\rm g}/\partial p_i$ are calculated by applying the product rule on the tree-level expression for the bispectrum in Eq.~\eqref{equ:galaxy-bispectrum} keeping only the terms with $P^{\rm nw}_m$ and $q$ fixed at fiducial cosmology where $q=1$. More specifically,
\begin{align}
    \frac{\partial B^{\rm w}_{\rm g}}{\partial p_i} & 
    \approx 
    2 \frac{1}{q^6} Z_2(\vec k_1, \vec k_2) Z_1(\vec k_1) Z_1 (\vec k_2') \left[P^{\rm nw}_{\rm m}(k_1|p_i^{\rm fid}) \frac{\partial P^{\rm w}_{\rm m}(k_2)}{\partial p_i} \right. \nonumber\\
    & \left. + P^{\rm nw}_{\rm m}(k_2|p_i^{\rm fid}) \frac{\partial P^{\rm w}_{\rm m}(k_1)}{\partial p_i}  \right] + {\rm 2~cyc.~perm.}
\end{align}
where
\begin{align}
    \frac{\partial P^{\rm w}_{\rm m}}{\partial p_i} = P_{\rm m}^{\rm nw} (k, \mu) \mathcal{D}(k, \mu)\frac{\partial O(k|\vec{p})}{\partial p_i}.
\end{align}

The parameter vector for the bispectrum is slightly different than that of the power spectrum:
\begin{align}
    \vec{p}=(N_{\rm eff}, \theta_\star, \omega_b, \omega_c, A_s, n_s, \tau, Y_p;\; \Vec{b}_1, \vec{b}_2, \vec{b}_{s^2};\; {a}_n, {b}_m),\nonumber
\end{align}
where we have the usual set of cosmological parameters, but now with two additional sets of second-order bias parameters $b_2$ and $b_{s^2}$. Note that we do not model the effects of the second-order bias parameters in the power spectrum, so this is not a consistent truncation in the expansion of $\delta_g$ in perturbation theory; however, we are adhering to taking the lowest order term in the power spectrum and bispectrum observables themselves. 

For the polynomial parameters that marginalize over systematics, we have $a_n$ and $b_m$ defined differently for the bispectrum than in the power spectrum (see Eq.~\eqref{equ:bs-poly-broadband} for more details). We choose for the fiducial set of parameters $b_{m\le 1}$ for the broadband version and $a_{n\le 4}, b_{ m\le 3}$ for the BAO wiggle version. The choice of polynomials and how they impact our results are discussed in Sec.~\ref{sec:forecast_dependencies}. 

{Note that the combined constraints from the power spectrum and the bispectrum are obtained by simply adding the corresponding Fisher matrices (P+B hereafter), while their covariances ($C^{PB}$) are ignored in this study (see the next paragraph for caveats). In this treatment, the polynomial coefficients in the power spectrum and the bispectrum Fisher matrices are treated as independent parameters, since we assume that they marginalize over systematic effects that affect these observables differently.}

{We caution the reader that this simplification only gives the lower bound of the real P+B constraint, and the missing $C^{PB}$ contribution can in principle be significant \citep{YankelevichPorciani2019}. 
{In reality, our P+B constraints are generally better than the P-only constraint and only slightly better than the B-only constraint (except for BOSS), which indicates that the joint constraint is dominated by the bispectrum. We expect therefore that the full results including $C^{PB}$ to fall somewhere in between the B-only and the P+B constraint, and will indicate our results for P+B in tables with a $>$ sign as a reminder of this fact.}
 } 

In order to evaluate the integral in Eq.~\eqref{equ:fisher-bispectrum}, we use a quasi Monte-Carlo method based on low-discrepancy sequences, more precisely, the Sobol sequence~\cite{sobol1967distribution}. Compared with integration on a regular grid or the ordinary Monte-Carlo integration method, the Sobol sequence method features a much faster convergence: $\mathcal{O}(N^{-1})$ versus $\mathcal{O}(N^{-0.5})$ where $N$ is the number of sampling points. For the 5-dimensional integral in the bispectrum Fisher matrix above, we only needed a total of $10^4 - 10^5$ sampling points for it to converge with a relative error below $5\%$.

Finally, we note also that since we are taking our derivatives using the chain rule on the tree-level expression for the bispectrum, where we make use of the power spectrum wiggle extractions, we do not perform a wiggle extraction directly on the bispectrum itself for calculating our Fisher forecast. 
In real data analysis, however, one would need to extract the wiggles directly from the measured bispectrum. This is best done by going into the $(k_1, \delta, \theta)$ coordinates. We show a naive attempt of directly using the same extraction algorithm for the power spectrum on the bispectrum in Appendix \ref{app:bao-extraction}. Since the algorithm assumes a  near-constant period in wiggles, we see a high-fidelity recovery of the BAO information for constructive configurations but worse performance for the destructive ones.

\subsection{Survey specifications}

We will forecast the constraints for a variety of galaxy redshift surveys: 
BOSS\footnote{\url{https://www.sdss.org/surveys/boss/}}, DESI\footnote{\url{https://www.desi.lbl.gov/}}, Euclid\footnote{\url{https://www.cosmos.esa.int/web/euclid/euclid-survey}}, PFS\footnote{\url{https://pfs.ipmu.jp/}}, SPHEREx\footnote{\url{https://spherex.caltech.edu/}; for the number density see \url{https://github.com/SPHEREx/Public-products/blob/master/galaxy_density_v28_base_cbe.txt}} and Roman Space Telescope\footnote{\url{https://roman.gsfc.nasa.gov/}}. Note that for SPHEREx, we do not use all five samples listed as in past forecasts~\cite{DoreBock2014}, but only the sample with the best redshift accuracy between $\sigma_z/(1+z) = 0$ and 0.003, amounting to negligible damping of modes along the line-of-sight due to photometric redshift errors for the scales we consider. For Roman, instead of using both the $H_{\alpha}$ and the $O_{\rm III}$ samples, we restrict only to the $H_{\alpha}$ sample which dominates at lower redshifts up to $z \approx 1.8$~\cite{EiflerMiyatake2021}.
For each survey, the key survey parameters include the mean galaxy number density at different redshift bins $\bar{n}_{\rm g}(z_i)$ and the sky coverage $f_{\rm sky}$, which are summarized in Table~\ref{tab:surveys}.

For comparison, we also include an idealized survey in the cosmic variance limit (CVL), setting $\bar{n}_{\rm g}= \infty$, $f_{\rm sky}=1$, and $z_{\rm max}=4$, while maintaining the BAO reconstruction rate fixed at 0.5.

All of our LSS results are combined with a CMB Fisher matrix for a mock Planck 2018 experiment with $\Lambda$CDM+$N_{\rm eff}$+$Y_{\rm p}$ following the formalism and specifications detailed in Ref.~\cite{BaumannGreen2018}. The Planck-only constraint gives $\sigma_{N_{\rm eff}}=0.32$, and serves as our baseline when comparing with CMB + LSS constraints in the next section.

\begin{figure}
    \centering
    \includegraphics[width=\linewidth]{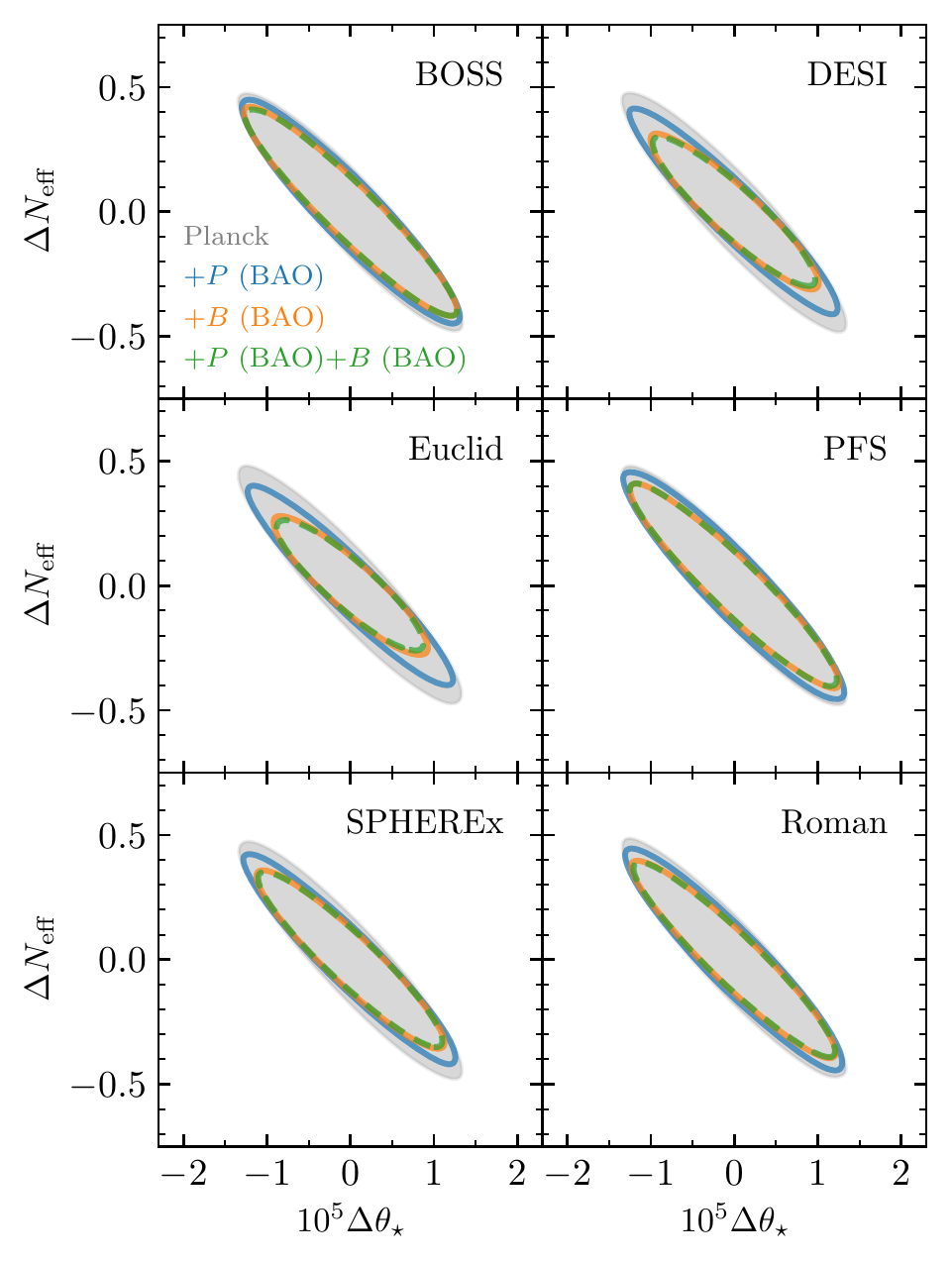}
    \caption{
    Marginalized 68\% confidence-level contours for $N_{\rm eff}$ and $\theta_\star$ for Planck alone (filled grey ellipses),  and Planck combined with various LSS surveys (colored lines) in our fiducial setup where the information comes from the BAO wiggle part of the LSS observables. Adding the power spectrum (P) to Planck (blue solid) improves on Planck alone, while adding the bispectrum alone (orange solid) to Planck represents an even further improvement. Planck + P + B (green dashed) does not do much better than Planck + B alone since the bispectrum is much more constraining than the power spectrum ({note that though correlations between P and B are not included in this study, this conclusion remains true}). Here we have marginalized over all the $\Lambda$CDM parameters, $Y_{\rm p}$, galaxy biases ($b_1$ for the power spectrum, and $\{b_1$, $b_2$, $b_{s^2}\}$ for the bispectrum), and polynomial coefficients for the effects of systematics.  }
    \label{fig:contours}
\end{figure}

\begin{table}
    \begin{subtable}{\linewidth}
        \begin{tabularx}{\textwidth}{XXXl}
            \toprule
            Planck+Survey & P & B & P+B \\
            \midrule
BOSS & 0.30 & 0.28 (0.30) & {$>$0.28 (0.29)} \\
DESI & 0.27 & 0.21 (0.26) & $>$0.20 (0.25) \\
Euclid & 0.26 & 0.18 (0.25) & $>$0.17 (0.24) \\
PFS & 0.30 & 0.27 (0.29) & $>$0.27 (0.29) \\
SPHEREx & 0.28 & 0.24 (0.27) & $>$0.23 (0.27) \\
Roman & 0.30 & 0.26 (0.29) & $>$0.26 (0.28) \\
CVL & 0.08 & 0.08 (0.15) & $>$0.06 (0.07) \\
            \bottomrule
        \end{tabularx}
        \caption{Our fiducial results of BAO-only constraints from power spectrum and bispectrum, and a Planck 2018 Fisher matrix are included for all cases. {Note that the P+B constraint does not have $C^{PB}$ in consideration, thus only indicates a lower bound.}}
        \label{tab:constraints-bao}
    \end{subtable}
    
    \begin{subtable}{\linewidth}
        \begin{tabularx}{\textwidth}{XXXl}
            \toprule
            Planck+Survey & P & B & P+B\\
            \midrule
BOSS & 0.23 & 0.22 (0.27) & $>$0.18 (0.22) \\
DESI & 0.13 & 0.11 (0.16) & $>$0.09 (0.12) \\
Euclid & 0.12 & 0.09 (0.14) & $>$0.08 (0.11) \\
PFS & 0.22 & 0.20 (0.27) & $>$0.17 (0.21) \\
SPHEREx & 0.16 & 0.14 (0.21) & $>$0.12 (0.15) \\
Roman & 0.19 & 0.17 (0.25) & $>$0.15 (0.19) \\
CVL & 0.05 & 0.04 (0.06) & $>$0.03 (0.04) \\
            \bottomrule
        \end{tabularx}
        \caption{Same as above, but for {total} constraints including the broadband and BAO wiggles. {Note that the P+B constraint does not have $C^{PB}$ in consideration, thus only indicates a lower bound.}}
        \label{tab:constraints-broadband}
    \end{subtable}
    
    \caption{Forecasted joint constraints on $N_{\rm eff}$ (68\% CL) from the Planck 2018 CMB experiment ($\Lambda$CDM+$N_{\rm eff}$+$Y_{\rm p}$) and different LSS surveys, after marginalizing over ($b_1, b_2, b_{s^2}$) and polynomial coefficients. For reference, the forecasted Planck-only constraint is $\sigma_{N_{\rm eff}}=0.32$. {We report constraints at the fiducial $k_{\rm max}=0.2\,h\,{\rm Mpc^{-1}}$, and also show in the parenthesis results with a more conservative assumption where the bispectrum has $k_{\rm max}=0.15\,h\,{\rm Mpc^{-1}}$. For joint CMB+P+B constraints, we only show the lower bound computed by ignoring the correlations between the power spectrum and bispectrum modes, {but do not expect the full results to vary much since they cannot be worse than the bispectrum-only results}.}
    }
    \label{tab:constraints}
\end{table}

\begin{figure*}
    \centering
    \includegraphics[width=\linewidth]{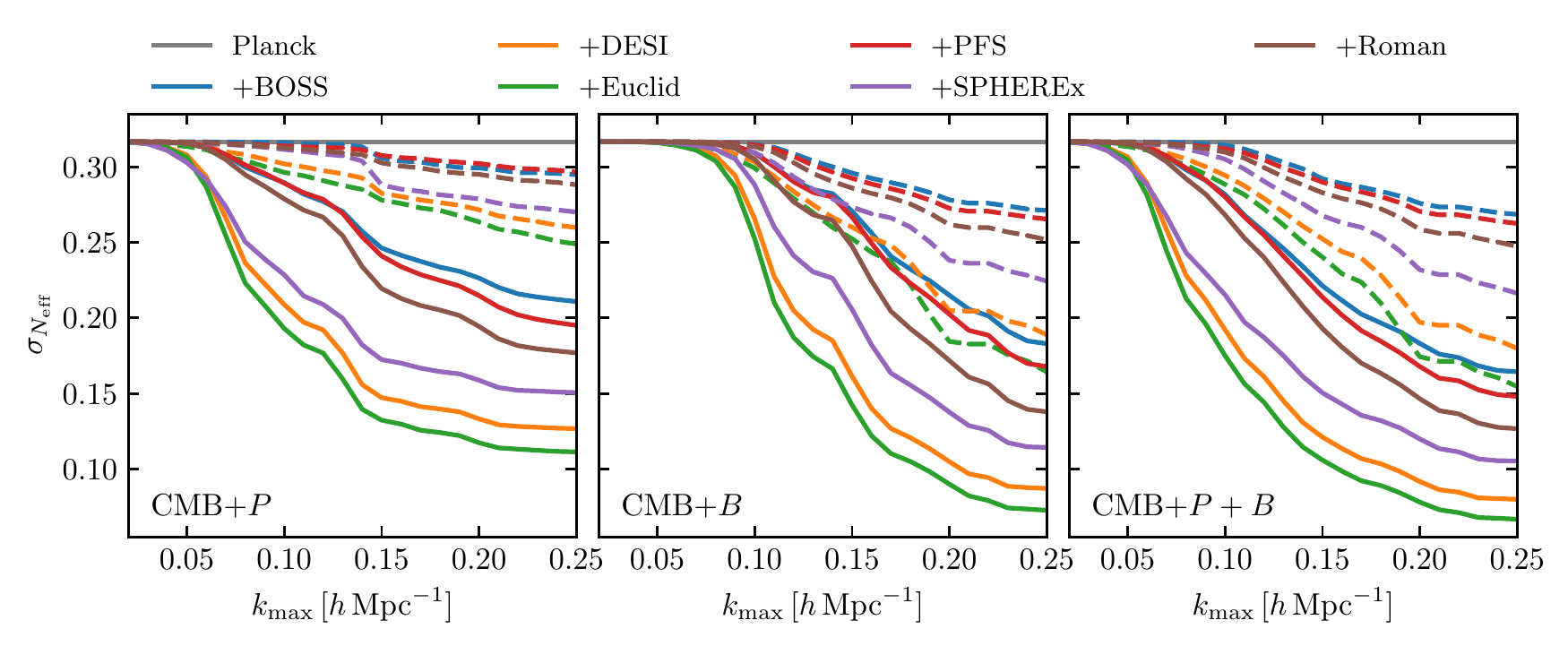}
    \caption{The 1D marginalized uncertainties for ${N_{\rm eff}}$ as a function of the maximum wavenumber $k_{\rm max}$ from Planck combined with the power spectrum (left), the bispectrum (middle) and P+B (right, {with $C^{PB}$ ignored}) for various LSS surveys. The fiducial constraints where the information comes from BAO wiggles only are shown in dashed lines, whereas the total constraints using the broadband and the BAO wiggles are shown in solid lines. As in Fig.~\ref{fig:contours}, these are marginalized results over $\Lambda$CDM parameters, $Y_p$, bias parameters and polynomial coefficients.
    }
    \label{fig:kmax_dependence}
\end{figure*}

\section{Results}
\label{sec:results}

We now present the constraints on $N_{\rm eff}$ from the power spectrum and the bispectrum for the various surveys using the Fisher formalism presented above.

\subsection{Fiducial results}
\label{sec:fiducial_results}

In Fig.~\ref{fig:contours}, we show our fiducial results for the constraints on $\theta_\star$ and $N_{\rm eff}$ using BAO wiggles for various LSS surveys. As in Ref.~\cite{BaumannGreen2018} for the case of the power spectrum, the LSS constraints by themselves are not as competitive as Planck alone. So here we show their joint constraints with Planck. The power spectrum (P), bispectrum (B), and joint P+B ({ignoring $C^{PB}$}) constraints with Planck are shown in red, green, and blue respectively, whereas Planck alone is plotted in grey. %


To start with, we compare the improvement of adding the LSS power spectrum to  the Planck-only constraints. Typically we see small improvements for most surveys except DESI and Euclid. This is consistent with that in Ref.~\cite{BaumannGreen2018}. Then for the LSS bispectrum, we can also see improvement upon the Planck-only result for different surveys, though for BOSS and PFS the improvement is not significant. For Euclid, we see that with bispectrum there is about a factor of $\sim 1.5$ improvement in $\sigma_{N_{\rm eff}}$ and $\theta_\star$ when compared to Planck-only. Finally, the combination P+B offers negligible further improvement upon bispectrum alone. 

%

More precisely, we show in Tab.~\ref{tab:constraints-bao} the 1D marginalized constraints for ${N_{\rm eff}}$ from BAO wiggles. Compared to the Planck constraint $\sigma_{N_{\rm eff}}=0.32$, the final CMB+P+B result is better by a factor of  ranging from 1.15 for BOSS to 1.84 for Euclid. We note that DESI and Euclid have the best improvements because of the large volume they probe.

Tab.~\ref{tab:constraints-broadband} shows similar results but for using both the broadband and the wiggle parts of the LSS observables. As expected, these results are better than using the BAO wiggles alone and would reflect the reality if all systematics could be successfully controlled to yield reliable broadband measurements. Here again, the improvement from the bispectrum alone over the power spectrum alone is about a factor of 1.05 -- 1.30, whereas going to joint P+B results improves upon the power spectrum only by about 1.23 -- 1.50. Not limiting ourselves to the BAO wiggles only, the best measurement from next-generation surveys would allow us to probe $N_{\rm eff}$ with $\sigma_{N_{\rm eff}}=0.08$, a factor of 4 improvement from the Planck alone result of 0.32, and merely a factor of 2 from an actual CVL experiment.  

As stated before, we also checked that the improvements from LSS power spectrum or bispectrum based on a CMB-Stage 4 experiment instead of Planck would be negligible. For a CMB-Stage 3 experiment, there is insignificant improvement from power spectrum BAO for all surveys as was also shown in Ref.~\cite{BaumannGreen2018}, but a slight improvement (by a factor of $\sim 1.1$) from the bispectrum BAO wiggles for DESI and Euclid.


In sum, we find that the LSS bispectrum signals, both the BAO wiggles and total, can help to improve the Planck-only constraint on $N_{\rm eff}$. The Planck+B also have a better constraint on $N_{\rm eff}$ than the Planck+P.

\subsection{Forecast Dependencies}
\label{sec:forecast_dependencies}

{The Fisher information is dependent on $k_{\rm max}$, and the constraint on $N_{\rm eff}$ is also affected by the polynomial parameters used for marginalizing over systematics, as well as other cosmological parameters that are beyond $\Lambda$CDM cosmology but are highly degenerated with $N_{\rm eff}$.} {We now investigate the dependencies of the Fisher matrix on these setup choices.}

\subsubsection{Varying $k_{\rm max}$}

Recall that throughout this work we set an upper limit $k\leq k_{\rm max}^{P}$ for the power spectrum and $k_1, k_2, k_3 \leq k_{\rm max}^{B}$ for the bispectrum Fisher computation, where we used the fiducial values $k_{\rm max}^{P}=0.2\,h\,{\rm Mpc}^{-1}$ for power spectrum BAO wiggles, $k_{\rm max}^{P}=0.5\,h\,{\rm Mpc}^{-1}$ for power spectrum broadband, and $k_{\rm max}^{B}=0.2\,h\,{\rm Mpc}^{-1}$ for both cases in the bispectrum. Now we explore the dependence of our forecast results for each survey in Fig.~\ref{fig:kmax_dependence} as we vary $k_{\rm max}$ from $k_{\rm min}$ to $0.25\, h\,{\rm Mpc}^{-1}$. In the left, middle and right panels, we show respectively the power spectrum, bispectrum and the joint P+B constraints. We do this for both the BAO-only (dashed lines) and the total
results (solid lines).

As expected, the constraining power gets better with higher $k_{\rm max}$. It does so faster for the bispectrum than for the power spectrum because the number of triangle configurations increases faster with $k_{\rm max}$ than the number of $k$-modes in the power spectrum. Additionally, for the power spectrum results, we see that the BAO constraints get a relatively sharp decrease near around $0.14\,h\,{\rm Mpc}^{-1}$ after the third peak in the BAO wiggles. For the bispectrum, this sharp decline is not at the same place as in the power spectrum, reflecting the fact that the information on $N_{\rm eff}$ is not coming from a peak in the oscillations with respect to one of the $k$'s, but rather in the  interference between two $k$'s (see Fig.~\ref{fig:bispectrum-bao-extraction}). %
Finally, this feature is carried over into the P+B BAO constraints since they are dominated by the bispectrum results for $k \gtrsim 0.08\,h\,{\rm Mpc}^{-1}$.

Now comparing between the surveys, Euclid gives the most optimistic result, for which the BAO constraints could reach $\sigma_{N_{\rm eff}}\approx 0.25$ for the PS and $\sigma_{N_{\rm eff}}\approx 0.17$ for the bispectrum (whereas the 
total
constraints reach $\sigma_{N_{\rm eff}}\approx 0.12$ for the PS and $\sigma_{N_{\rm eff}}\approx 0.07$ for the bispectrum) for our fiducial $k_{\rm max}=0.2 h\,{\rm Mpc}^{-1}$. The results become even better at higher $k_{\rm max}$. We caution the readers however that past $k \sim 0.2 h\,{\rm Mpc}^{-1}$, the linear scale modeling that we use for both the power spetrum and the bispectrum becomes less accurate. %

\begin{figure}
    \centering
    \includegraphics[width=\linewidth]{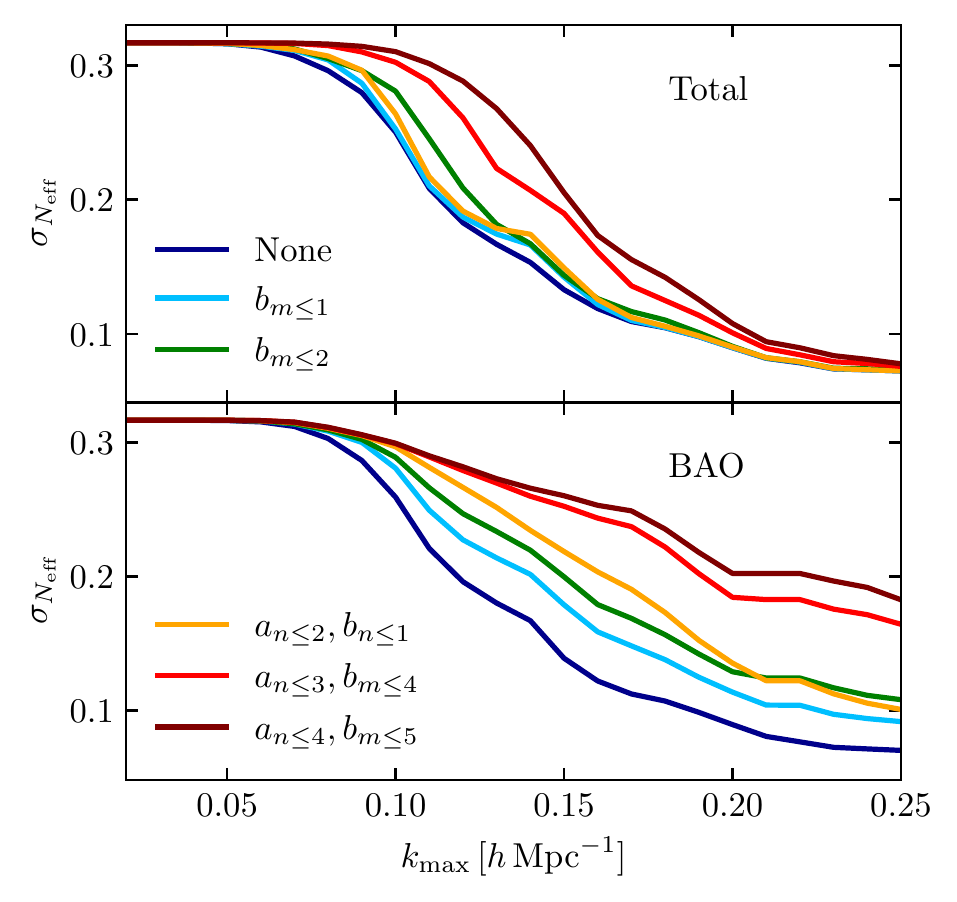}
    \caption{The 1D marginalized uncertainties for $N_{\rm eff}$ from Planck combined with the Euclid bispectrum for various choices of polynomial terms that are used to marginalize over systematics. We show in the top panel the total constraints from the broadband bispectrum and BAO wiggles, and in the bottom panel the constraints from BAO wiggles alone, where the polynomials are defined differently (see Eqs.~\ref{equ:bs-poly-broadband} and ~\ref{equ:bs-poly-wiggle}). The fiducial set of polynomial terms are
    $b_{m\le 1}$ for the broadband (light blue) and $a_{n\le3},~b_{m\le 4}$ for BAO-only (red).
    }
    \label{fig:bs-poly}
\end{figure}

\subsubsection{Varying the polynomial model}

As we introduced in the previous section, the polynomial terms are included in the galaxy power spectrum and bispectrum modeling to account for uncertainties say, in measurement or modeling measurement, or in the extraction the BAO wiggles. However, the specific number of terms we choose to include could impact the forecasted constraints significantly. Including not enough parameters, one may get too optimistic forecast; including too many parameters could over-penalize the analysis. Recall that we follow Ref.~\cite{BaumannGreen2018} to choose a specific set of terms for the power spectrum and keep the same powers of $k$ for the bispectrum, namely, $b_{m\le 1}$ for the total bispectrum 
and $a_{n\le 3},b_{ m\le 4}$ for BAO wiggles.

In Fig.~\ref{fig:bs-poly} we vary the fiducial set of polynomial parameters for the Euclid bispectrum forecast. 
For the total in the upper panel,
 we see that the impact of polynomial terms is not as significant when using a high $k_{\rm max}$, but the additive coefficients $b_m$ do cast significant impact at lower $k_{\rm max}$. For $k_{\rm max}^{B} = 0.2\,h\,{\rm Mpc}^{-1}$, our fiducial choice of $b_{m\le 1}$ (red) does not deviate much from other choices.

For the BAO wiggles in the bottom panel, we see that the polynomial coefficients do affect the result more significantly. At $k_{\rm max} = 0.2\,h\,{\rm Mpc}^{-1}$, the constraints on $N_{\rm eff}$ can vary between $\sim 0.1$ to $\sim 0.2$ depending on the choice of polynomials. Our fiducial setup of $a_{n\le 3},b_{m\le4}$ (red) is on the more conservative side.

\begin{figure}
    \centering
    \includegraphics[width=\linewidth]{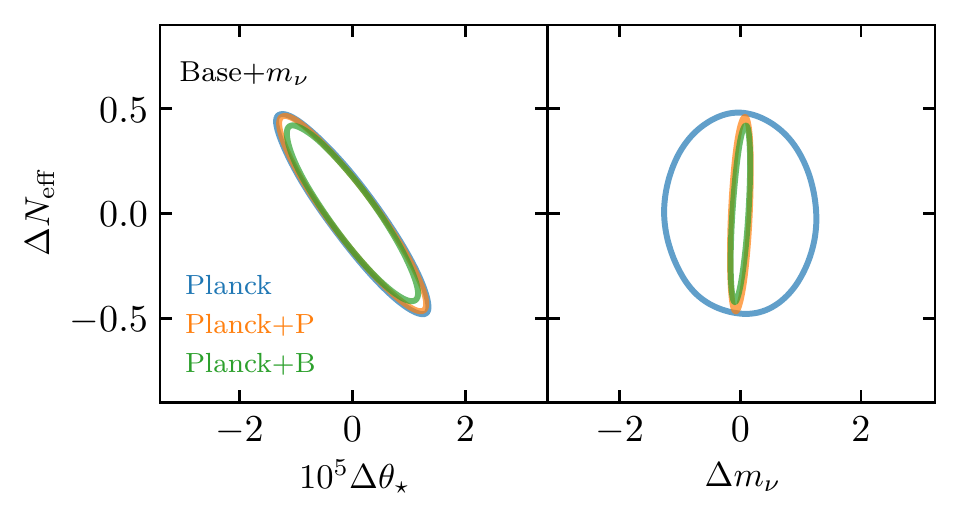}
    \includegraphics[width=\linewidth]{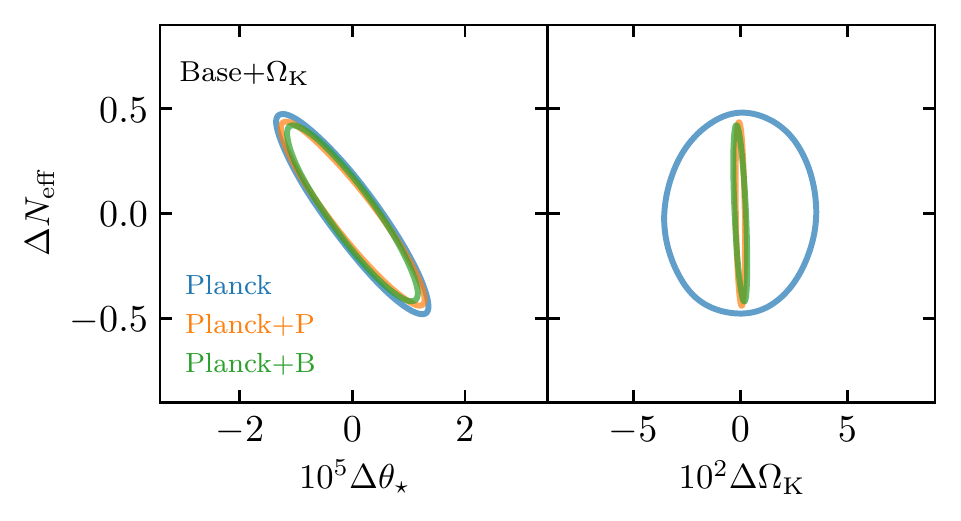}
    \includegraphics[width=\linewidth]{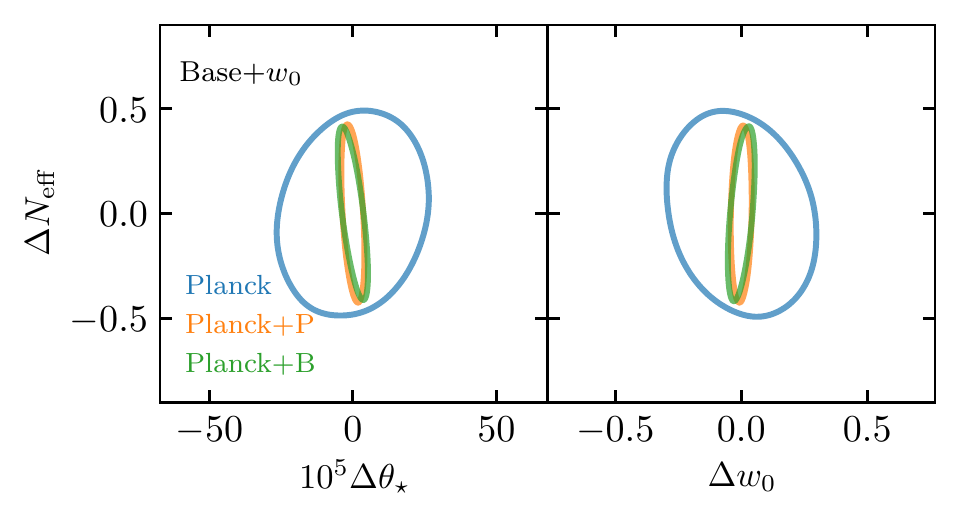}
    \caption{{The impact of marginalizing over additional parameters beyond $\Lambda$CDM. In each row, we add a new parameter $m_\nu$, $\Omega_{\rm K}$, or $w_0$ to the baseline 
    Planck and DESI (BAO only) forecasts. We show 68\% CL contours of $N_{\rm eff}$ and $\theta_\star$ in the left panel of each row, and those of $N_{\rm eff}$ and the new parameter in the right one. After adding these parameters, there is a significant degradation in both the power spectrum (P) or bispectrum (B) BAO wiggle-only constraints. In particular, with $w_0$ added, we also see that the constraint on $\theta_\star$ degrades significantly due to the high degeneracy between $\theta_\star$ and $w_0$.}
    }
    \label{fig:additional_parameters}
\end{figure}

\begin{table}
        \begin{tabularx}{\linewidth}{XXXl}
            \toprule
            Parameters & Planck & Planck+P & Planck+B \\
            \midrule
            Base & 0.32 & 0.27 & 0.21 \\
            Base+$m_\nu$ & 0.32 & 0.31 & 0.27 \\
            Base+$\Omega_{\rm K}$ & 0.32 & 0.29 & 0.28 \\
            Base+$w_0$ & 0.33 & 0.28 & 0.27 \\
            \bottomrule
        \end{tabularx}
        \caption{{Constraints on $N_{\rm eff}$ (68\% CL) after marginalizing over one extended parameter, for the Planck and DESI (BAO only) configurations. Here ``Base'' denotes the default set of parameters listed in Sec.~\ref{sec:fisher}. {The improvement of the bispectrum over the power spectrum seen in the base model starts to become erased when considering the $w$CDM model or a model with curvature.}
        }
        }
        \label{tab:additional_params}
    \end{table}

\subsubsection{Additional cosmological parameters beyond $\Lambda$CDM}

{Besides the standard parameter space we explored ($N_{\rm eff}, \theta_\star, \omega_b, \omega_c, A_s, n_s, \tau, Y_p$), we also investigate the known degeneracy between $N_{\rm eff}$ and a set of extended parameters -- the neutrino mass $m_{\nu}$, the curvature $\Omega_{\rm K}$, and the dark energy equation-of-state parameter $w_0$ ($w_0 \equiv P_{\rm DE}/\rho_{\rm DE}$) -- and how they could affect our forecasts. As a test case, we evaluate the impact of these new parameters on the Planck+DESI BAO-only constraints, by varying the parameters one at a time from their default values $m_{\nu}=0.06\,\rm eV$, $\Omega_{\rm K}=0$, and $w_0=-1$.}

{In Tab.~\ref{tab:additional_params}, we show how the Planck and Planck+DESI constraints on $N_{\rm eff}$ degrade after marginalizing over each of the extended parameters. Typically, we find that the Planck-only constraints on $N_{\rm eff}$ are similar to before, about $\sim 0.32$, while there is a degradation for the Planck+P and Planck+B constraints. 
Specifically,
the Planck+B constraints typically reach $\sigma_{N_{\rm eff}} \sim 0.27$ after 
marginalizing over
these new parameters, which is about $30\%$ worse than the fiducial result of $\sigma_{N_{\rm eff}} = 0.21$.}

{In Fig.~\ref{fig:additional_parameters}, we visualize the degeneracy between $N_{\rm eff}$ and each extended parameter ($m_\nu$, $\Omega_{\rm K}$, and $w_0$ from top to bottom). We compare the 68\% CL contours of the Planck-only, Planck+P, and Planck+B constraints in the $(\theta_\star, N_{\rm eff})$ plane in the left column, and in the (X, $N_{\rm eff}$) plane in the right column, where X is the new parameter. An improvement in $\sigma_{N_{\rm eff}}$ for both the Planck+P and Planck+B constraints upon the Planck-only constraint still remains. However, it is not as significant as before (see Tab.~\ref{tab:additional_params}). Perhaps the most striking is the improvement of the constraints on the extended parameters themselves ($m_\nu$, $\Omega_{\rm K}$, and $w_0$) when adding LSS data to Planck, as shown in the right panels of Fig.~\ref{fig:additional_parameters}. With further inspection, we find more Fisher information of $m_\nu$ and $\Omega_{\rm K}$ in the LSS than the CMB (Planck), explaining the improvement in constraints of these parameters. For $w_0$, the improvement is primarily due to the fact that LSS breaks the strong degeneracy between $w_0$ and $\theta_\star$ in the CMB. 
}

{There is also an interplay between these new parameters and other fiducial parameters ($\omega_{\rm b}$, $\omega_{\rm c}$, etc.). For example, in Fig.~\ref{fig:additional_parameters} we find that the constraint on $\theta_{\star}$ degrades significantly after adding $w_0$, and this is due to the very strong degeneracy between $w_0$ and $\theta_\star$, though the impact on $N_{\rm eff}$ is much less significant.}

\begin{figure}
    \centering
    \includegraphics[width=\linewidth]{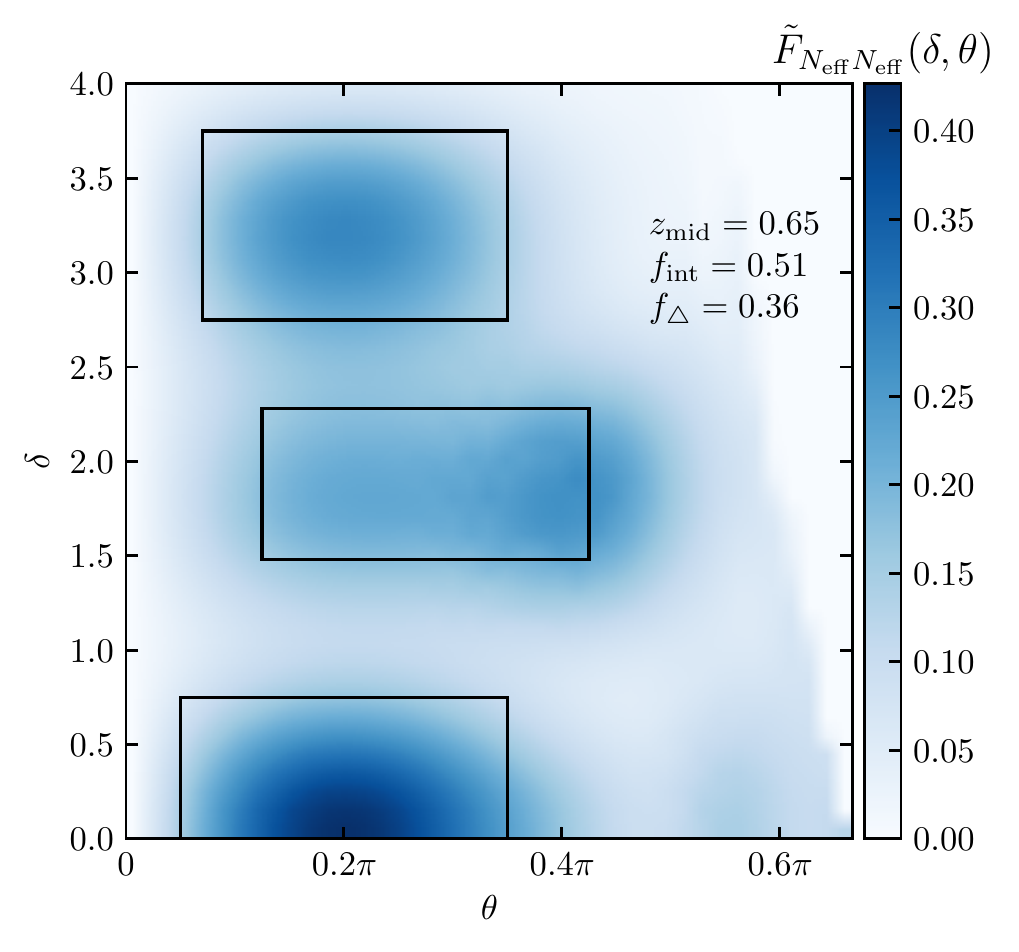}
    \caption{The normalized contribution $\tilde{F}_{N_{\rm eff}N_{\rm eff}}(\delta, \theta)$ to the Fisher information for the parameter $N_{\rm eff}$ as a function of Child18 coordinates $\delta$ and $\theta$. We integrate the integrand of the Fisher matrix element $F_{N_{\rm eff}N_{\rm eff}}$ (Eq.~\ref{equ:fisher-bispectrum}) over $k_1$ as well as the angular variables ($\mu_1$, $\mu_2$) and normalize over the whole plane (see Eq.~\ref{eq:F_Neff_Neff}). 
    This is an example using the first redshift bin ($0.6\le z\le0.7$) of the Euclid survey. The Fisher information shows peaks near even $\delta$'s which are the constructive interference configurations. Using only the boxed regions which represent about $\sim 36\%$ of the total  number of triangle configurations, we recover $\sim 51\%$ of the total Fisher information on $N_{\rm eff}$. Note that there are negligible information around $\delta = 4$ because there are no triangle configurations there given our conditions $0.01\,h\,{\rm Mpc}^{-1} \le k_1, k_2, k_3\le 0.2\,h\,{\rm Mpc}^{-1}$.
    }
    \label{fig:integrand}
\end{figure}

\subsection{Note on bispectrum interference and computational costs}

\label{sec:notes_on_interference}

One of the advantages of using bispectrum for spectroscopic survey is that one can exploit the interference structure in order to simplify the analysis. Most of the signal will be concentrated around the constructive configurations as explained in Sec.~\ref{sec:interference}. In this section we will investigate how much Fisher information we can preserve when choosing to measure only a subset of configurations.

In Fig.~\ref{fig:integrand} we present the Fisher information of $N_{\rm eff}$, $\tilde{F}_{N_{\rm eff} N_{\rm eff}}$, in the $(\delta, \theta)$ plane. Here $\tilde{F}_{N_{\rm eff} N_{\rm eff}}(\delta, \theta)$ is obtained from the same integrand as in Eq.~\eqref{equ:fisher-bispectrum}, but integrated over $0.01\,h\,{\rm Mpc}^{-1}\le k_1\le0.2\,h\,{\rm Mpc}^{-1}$ and $-1\le \mu_1, \mu_2\le 1$, and then normalized over the whole plane: 
\begin{align}
    \tilde{F}_{N_{\rm eff} N_{\rm eff}} & \propto  \int_{k_{\rm min}}^{k_{\rm max}} \dd k_1 \int_{-1}^{1}\dd \mu_1 \int_{-1}^1 \dd \mu_2 \left(\frac{\partial B }{\partial N_{\rm eff}}\right)^2 \nonumber \\ &  \frac{V k_1 k_2 k_3 \gamma(\cos \theta)\Sigma (\mu_1, \mu_2, \cos \theta)   }{8 \pi^4 s_{123} P(\vec k_1)P(\vec k_2)P(\vec k_3)} \nonumber\\ & \Theta(k_{\max}-k_2)\Theta(k_{\max}-k_3).
\label{eq:F_Neff_Neff}
\end{align}
Here the step function $\Theta$ ensures that $k_2, k_3 \le k_{\rm max}$. Thus $\tilde{F}_{N_{\rm eff} N_{\rm eff}}$ is the density distribution function of the bispectrum BAO information
in terms of $N_{\rm eff}$. 

As expected the information is mostly concentrated around constructive interferences $\delta = 0$ and 2. However, there are some deviations around $\delta = 4$. We have verified that this is due to imposing an upper bound $k_3 \leq k_{\rm max}$, which excludes some triangle configurations around $\delta = 4$. More specifically, since $\delta = 1$ corresponds to $k_2 - k_1 \approx 0.03\,h\,{\rm Mpc}^{-1}$, at $\delta = 4$ and for $k_1 \gtrsim 0.08\,h\,{\rm Mpc}^{-1}$ (a large part of the $k_1$ range noted above), we have that $k_2 \gtrsim k_{\rm max} = 0.2 h\,{\rm Mpc}^{-1}$ for which there are no triangle configurations available. 

Now, to quantify how the interference can help reduce computational costs without much loss of information, we first select the regions where most of the information is contained (see boxed regions in Fig.~\ref{fig:integrand}). Then we calculate the number of triangle configurations as well as the Fisher information enclosed in these regions, and report their fractions compared to the total. 

For a setup mimicking the first redshift bin ($0.6 \leq z \leq 0.7$) of the Euclid experiment, we get that 51\% of the Fisher information is enclosed within the boxed regions which contain 36\% of the triangle configurations. This reduction in the number of triangles can represent a significant cut in computation time during real data analysis: Cutting the data vector dimension by a factor $f_\triangle$ means a similar cut on time spent on computing the estimator and theory prediction in a MCMC analysis, as well as a significant larger cut ($\sim f_\triangle^2$) on the number of simulations required to generate the bispectrum covariance matrix.

Finally, we recommend doing the Fisher analysis before the real data analysis to identify the ideal boundaries for the regions with most information, as they could change depending on the survey setup. For example, we find that the peaked regions may move in the $\theta$ direction between different redshift bins. In the setup we explored, the peaked regions shift toward the right in the $\theta-\delta$ plane with higher redshift because of the redshift evolution of the galaxy bias and the linear growth rate.

\section{Discussion and Conclusion}
\label{sec:conclusions}

In this paper we forecast constraints on the effective number of neutrino species $N_{\rm eff}$ from various LSS surveys using the Fisher formalism, examining for the first time the impact of including bispectrum measurements with BAO wiggles. We present two versions of the forecasts where the information comes from the BAO wiggles alone, which is our fiducial results, as well as from the total bispectrum including both the broadband shape of the bispectrum and the BAO wiggles. %

We find for both cases that, although the LSS constraints alone are not competitive with Planck, combining the LSS constraints with Planck provides a clear improvement over Planck alone ($\sigma_{\rm N_{\rm eff}}=0.32$). This is in alignment with what the authors of Ref.~\cite{BaumannGreen2018} found for the power spectrum, which we also reproduce. Using BAO wiggles only, we find that Planck+B clearly improves upon Planck alone, with a $\sigma_{\rm N_{\rm eff}}$  ranging from 10\% to 40\% improvement depending on the survey. There is also a notable improvement from Planck+P to Planck+B of about 5\% - 30\% depending on the survey. Planck+P+B {(by ignoring the correlation between power spectrum and bispectrum modes $C^{PB}$, which is a lower bound of this constraint)} does not in general provides better constraints than Planck+B because the bispectrum constraints were already very good, except for the case of CVL where combining all data allows one to reach $\sigma_{\rm eff} = 0.06$.

When using the {total} bispectrum including both the broadband and wiggles, we obtain better constraining power as expected. The broadband information is valuable if the systematics in the measurement can be reliably controlled. Here the Planck+B constraint reaches $\sigma_{N_{\rm eff}}=0.09$ for Euclid, and as low as $\sigma_{N_{\rm eff}}=0.04$ for a CVL experiment up to $z_{\rm max} = 4$. However, measurements of the broadband are challenged by systematics and modeling uncertainties. The latter could be well-controlled using an effective field theory of LSS~\cite{BaldaufMercolli2015}, especially when using higher $k_{\rm max}$ than our fiducial choice of $k_{\rm max}=0.2\, {h\, {\rm Mpc}^{-1}}$.

{We caution the reader that some extensions of the $\Lambda$CDM model can weaken the improvement we have been reporting in the $\Lambda$CDM model from the LSS bispectrum, due to the strong correlations between the extended parameters and $N_{\rm eff}$. In particular, for the same CMB+LSS configuration, we found a $\sim 30\%$ degradation in the case of marginalizing over the parameters $m_\nu$, $\Omega_{\rm K}$, and $w_0$ one at a time. Simultaneously, we find the CMB+LSS data can significantly help improve the constraints on non-standard parameters like $m_\nu$, $\Omega_{\rm K}$, and $w_0$.} 
We also utilize the template modeled in Ref.~\cite{BaumannGreen2018} to study the constraints from the BAO phase shift. Similarly here, we see better performance for the bispectrum over the power spectrum. However, the phase-shift constraint is not as competitive as the BAO-only or total constraints. For example, for CVL the phase shift constraint with prior from Planck is $\sigma_{N_{\rm eff}}=0.27$ for the bispectrum. {This probe can however be useful for probing physical effects that mainly show up as a phase shift, such as the isocurvature perturbations}. 

%

%

Note that we have chosen $k_{\rm max}=0.2\, {h\, {\rm Mpc}^{-1}}$ {for the bispectrum forecasts}, 
the regime of validity for the tree-level bispectrum and linear theory. To push $k_{\rm max}$ higher into the weakly nonlinear regime, one may choose to add higher loop terms~\cite{LazanuLiguori2018}; alternatively, one may use the tree-level form with a nonlinear matter power spectrum and an effective second-order kernel $F_{2, \rm eff}$ fit from simulations in the nonlinear regime for the broadband modeling~\cite{ScoccimarroCouchman2001}. 

To fully simulate the wiggles and the broadband in the nonlinear regime, one would measure the bispectrum directly from simulations, though it takes special simulations capable of capturing the neutrino-induced BAO {effects}
However, this may not be necessary since, due to nonlinear damping of BAO signals at smaller scales, the BAO information will be limited there and there may not be significant improvement by extending $k_{\rm max}$ in the BAO modeling. However, there would be more information to be gained from the broadband in principle. EFTofLSS has been shown to be promising in this regard, and it may be worth applying it to the measurement of $N_{\rm eff}$~\cite{BaldaufMercolli2015}.

Additionally, we extended the concept of bispectrum interference, first explored in Ref.~\cite{ChildTakada2018} for the sound horizon measurement, by applying it to $N_{\rm eff}$ here. The bispectrum interference technique allows us to reduce computational cost by identifying the set of triangle configurations that contain the most information (those exhibiting constructive interference). We find for example that using only about a third of the triangles would give us half of the Fisher information in $N_{\rm eff}$ for the Euclid experiment's lowest redshift bin. This can dramatically reduce the computational challenges involved in measuring the bispectrum, especially when deriving covariance matrices from simulations.

The bispectrum interference coordinates $(k_1, \delta, \theta)$ also offer a natural way to extract the BAO wiggles from a bispectrum measurement. We attempted a naive application of the current wiggle extraction algorithm designed for the power spectrum directly on the bispectrum in Appendix~\ref{app:bao-extraction}. We found that the while the algorithm does not work well for the destructive interference configurations, these are exactly the configurations that do not contain much information on $N_{\rm eff}$ and can therefore be neglected in a real data analysis. 

We also showed that improvements are needed in the current algorithm are for getting better results in the constructive configurations in the bispectrum: While the main shape of the damping envelope is well captured for several $\delta = 0$ and $\delta = 2$ configurations tested, the amplitude of the first two peaks are not accurately measured. It may be  that the periodicity assumptions in the algorithm is a less good approximation in the case of bispectrum interference case than in the power spectrum. Therefore, improvements on the algorithm or a rigorous characterization of the errors induced are needed in order to directly measure BAO wiggles in the bispectrum. In our forecast, we have chosen to marginalize over a set of polynomials in order to capture some of the measurement errors induced. 


In sum, the next-generation LSS surveys can improve on current constraints on $N_{\rm eff}$ from Planck by up to a factor of 2 using observations in the linear regime. Future work could include extending the modeling of the galaxy bispectrum into the weakly non-linear regime, which may provide improvement upon even CMB-Stage 4 like experiments without requiring a more futuristic LSS survey. Developing wiggle extraction algorithms specifically tailored for the bispectrum interference could also open doors for alternative measurements the BAO wiggles in the bispectrum, which is useful for constraining physical effects that affect the BAOs such as $N_{\rm eff}$ and isocurvature perturbations. 


%
%

%

%

\begin{acknowledgments}
    This work was partially supported by NASA grant 15-WFIRST15-0008 Cosmology with the High Latitude Survey Roman Science Investigation Team. Part of this work was done at Jet Propulsion Laboratory, California Institute of Technology, under a contract with the National Aeronautics and Space Administration. We also acknowledge support from the SPHEREx project under a contract from the NASA/GODDARD Space Flight Center to the California Institute of Technology.
    The code and data sets used in this article are available at \url{https://github.com/yanlongastro/galaxy_survey}.
\end{acknowledgments}

\begin{appendix}

\begin{figure}[!htbp]
    \centering
    \includegraphics[width=\linewidth]{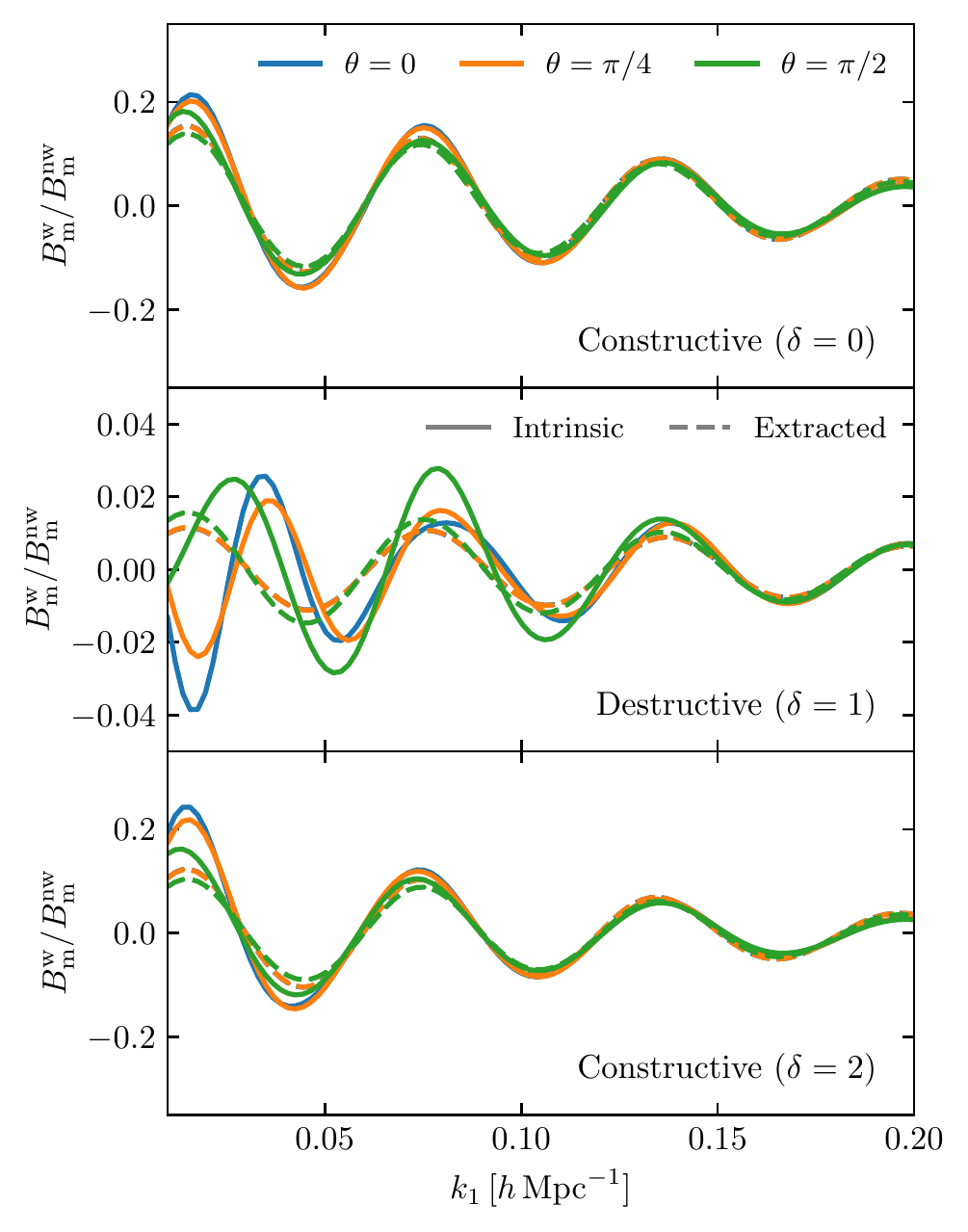}
    \caption{BAO wiggles in the bispectrum. We show the results for constructive interference ($\delta=0,2$) and destructive interference ($\delta=1$) configurations with $\theta = 0, \pi/4$ and $\pi/2$. The solid lines are intrinsic wiggles computed with the tree-level bispectrum where the wiggle and non-wiggle split has been performed on the theory prediction for the linear power spectrum using the standard wiggle extraction algorithm, while the dashed lines are wiggles directly extracted from the bispectrum $B_{\rm m}(k_1, \delta, \theta)$ using the same algorithm. The assumption of periodic oscillations in the algorithm breaks down more severely for the destructive interference, but these configurations can be ignored since they do not contain much information on $N_{\rm eff}$.}
    \label{fig:bispectrum-bao-extraction}
\end{figure}

\section{Extraction of BAO wiggles from bispectrum}
\label{app:bao-extraction}

To extract the BAO wiggles from the measured bispectrum, one would need to transform the bispectrum from the $(k_1, k_2, k_3)$ coordinates to the $(k_1, \delta, \theta)$ coordinates. 
{There the interference of wiggles are made explicitly manifest and wiggle extraction algorithms similar to those used in the power spectrum can be developed to extract them.}
In Fig.~\ref{fig:bispectrum-bao-extraction}, we show the results directly applying one of the standard algorithms developed for the power spectrum (that described in Appendix C of Ref.~\cite{BaumannGreen2018}) on a variety of $(\delta, \theta)$ configurations. In the top and bottom panels, we show the result for the constructive interference configurations $\delta = 0$ and 2, whereas in the middle panel we show the result for the destructive interference with $\delta = 1$, where the amplitude of the wiggles are about 10 times lower.

The solid lines, call it intrinsic wiggles, correspond to the ratio $B_{\rm m}^{\rm w}/B_{\rm m}^{\rm nw}$ where the wiggle and non-wiggle parts of the bispectrum are obtaining by computing the tree-level expressions with $P_{\rm m}^{\rm w}$ and $P_{\rm m}^{\rm nw}$, where the wiggle-extraction algorithm has been applied on the power spectrum. The dashed lines, call it extracted wiggles, correspond to applying the extraction algorithm directly on the matter bispectrum with wiggles, which is more in line with what one would do in an actual measurement when we do not make use of the measured power spectrum. These two methods end up defining a different non-wiggle bispectrum, hence the differences in the plotted ratios $B_{\rm m}^{\rm w}/B_{\rm m}^{\rm nw}$. The differences seen may be related to the fact that the extraction algorithm employed assumes a near-constant period in wiggles. This assumption is broken in slightly different ways for the wiggles in the bispectrum and in the power spectrum. 

It is clear from the plot that the difference is less significant for the constructive configurations ($\delta=0,2$) than for the destructive configurations ($\delta = 1$). Since one may select to only measure the constructive configurations where most of the information resides, this makes it possible to model the bispectrum wiggles by doing the split in the power spectrum space first and then using the tree-level expression to get the bispectrum wiggle predictions during an MCMC analysis, which would be faster than performing the split on every bispectrum configuration in the theory calculation. How well this works in practice will depend on the details of the algorithm chosen, and should be tested on a simulated data first prior to use in an actual data analysis.

In sum, developing extraction algorithm specifically tailored to the bispectrum interference, perhaps without the stringent assumption of periodicity could be useful. Alternatively, characterizing the error induced when applying a non-ideal extraction algorithm could be useful as well. Finally, this illustration is done on the matter bispectrum. Further tests with the galaxy bispectrum including the realism of redshift space distortions would be necessary as well.


%

%

\begin{figure}[!htbp]
    \centering
    \includegraphics[width=.9\linewidth]{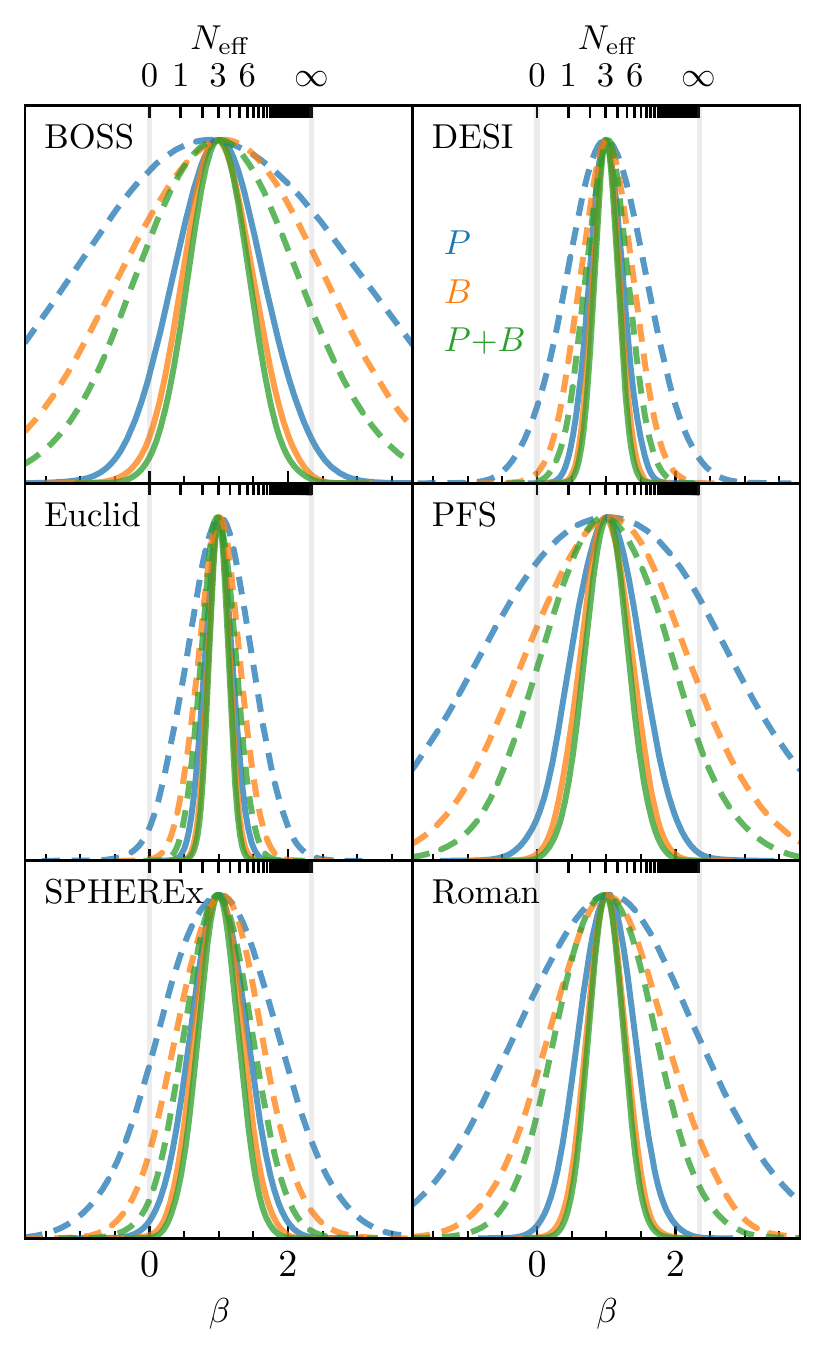}
    \caption{Forecasted posterior distribution for $\beta$ from measurements of phase shifts in various LSS survey. We show the distribution for the LSS power spectrum alone (blue), the bispectrum alone (orange), or the combined power spectrum + bispectrum (green). No CMB constraint on $N_{\rm eff}$ is added, but we explore including (solid) and not including (dashed) a Planck 2018 prior on the BAO scale $\alpha$ from the $\Lambda$CDM+$N_{\rm eff}$ cosmology. 
    }
    \label{fig:dist-beta}
\end{figure}

\section{Gaussian bispectrum covariance}
\label{app:N_tri}

To evaluate the Fisher matrix for the bispectrum, we start with the covariance matrix, e.g., the correlation between $B(\vec{k}_1, \vec{k_2}, \vec{k}_3)$ (for brevity B hereafter) and $B(\vec{k}_1', \vec{k}_2', \vec{k}_3')$ (for brevity $B'$). We consider the small range of $(\Vec{k}_i-\dd \vec{k}_i/2, \Vec{k}_i + \dd \Vec{k}_i/2)$ ($i=1, 2, 3$), the Gaussian contribution to the covariance matrix is given by~\cite{ChanBlot2017,YankelevichPorciani2019}
\begin{align}
    {\rm Cov}(B, B') & = \frac{V}{N_{\rm tri}} s_{123}  P_{\mathrm{obs}}\left(\vec k_{1}\right) P_{\mathrm{obs}}\left(\vec k_{2}\right) P_{\mathrm{obs}}\left(\vec k_{3}\right) \nonumber\\
    & \times \delta_{\rm D} (\vec{k}_1+\vec{k}_1') \delta_{\rm D} (\vec{k}_2+\vec{k}_2')\delta_{\rm D} (\vec{k}_3+\vec{k}_3').
\end{align}
Here $N_{\rm tri}$ is the number of modes, which follows $N_{\rm tri} = V_{123}/k_{\rm f}^6$,  where $k_{\rm f}^3=(2\pi)^3/V = V_{\rm f}$ is the fundamental volume and $V_{\rm 123}$ is the volume constrained by $[\vec k_i-\dd \vec k_i/2, \vec k_i + \dd \vec k_i/2]$ ($i=1,2,3$). Since $\vec{k}_1+\vec{k}_2+\vec{k}_3=0$, we only need $\vec k_1$ and $\vec{k}_2$.

To derive $V_{123}$, we may decomposed the $\vec{k}$'s into spherical coordinate, e.g., $(k_1, \mu_1, \phi_1)$ and $(k_2, \mu_2, \phi_2)$, where $\phi_1$ and $\phi_2$ are azimuthal angles. Note that in RSD there is azimuthal symmetry for the triangle configuration, we may set $\phi_1=0$ without loss of generality. By definition we have
\begin{align}
    V_{123} & = \int_{[\vec{k}_{1, \pm}]} \dd^3 \vec{p} \int_{[\vec{k}_{2, \pm}]} \dd^3 \vec{q} \int_{\infty} \dd^3 \vec{r} \delta^{\rm D} (\vec{p}+\vec{q}+\vec{r}) \nonumber\\
    & = 2 \pi k_1^2 \dd k_1 \dd \mu_1 \cdot k_2^2 \dd k_2 \dd \mu_2 \dd \phi_2.
\end{align}
Here $[\vec{k}_{i, \pm}]$ denotes a volume region constrained by $[\vec k_i-\dd \vec k_i/2, \vec k_i + \dd \vec k_i/2]$. To relate $\phi_2$ to $k_3$, we have 
\begin{align}
    k_3^2 & = k_1^2+k_2^2 + 2k_1k_2\cos \theta;\\
    \cos \theta & = \sqrt{1-\mu_1^2}\sqrt{1-\mu_2^2} \cos \phi_2 + \mu_1 \mu_2; \\
    \frac{\partial \phi_2}{\partial k_3} & = -\frac{2\pi k_3}{k_1 k_2} \cdot \Sigma(\mu_1, \mu_2, \cos \theta),
\end{align}
where (also refer Ref.~\cite{YankelevichPorciani2019})
\begin{align}
    \Sigma = \frac{1}{2 \pi \sqrt{1-\cos^2 \theta-\mu_{1}^{2}-\mu_{2}^{2}+2 \mu_{1} \mu_{2}\cos\theta}}.
\end{align}
One may check that $\int_0^1 \dd \mu_1 \int_0^1 \dd \mu_2 \Sigma =1$. Numerically, we will encounter singularities and waste many sampling points if choosing $\mu_1$ and $\mu_2$ as our angular coordinate, especially when $\abs{\cos \theta} \to 1 $. This is also illustrated in Fig. 1 of Ref.~\cite{YankelevichPorciani2019} (see the lower two panels, where there are no triangle configurations outside the ellipse). To avoid this problem, we can perform a coordinate transformation and use new coordinate $(\mu_s, \zeta)$ such that
\begin{align}
    \mu_{1} \to \tilde{\mu}_{1} & =\frac{\sqrt{2} \left(\mu_{1}-\mu_{2}\right)}{2 \sqrt{1-\cos \theta }} = \cos \zeta \sqrt{1-\mu_s^2};\\
    \mu_{2} \to \tilde{\mu}_{2} & =\frac{\sqrt{2} \left(\mu_{1}+\mu_{2}\right) }{2 \sqrt{1+\cos \theta }} = \sin \zeta \sqrt{1-\mu_s^2}; \\
    \Sigma \dd \mu_1 \dd \mu_2  & \to \frac{1}{2\pi} \dd \mu_s \dd \zeta.
\end{align}
The setup can dramatically improve efficiency and accuracy of the integration.

We also note that for $\theta$ and $2\pi - \theta$ (or equivalently $\phi_2 \to 2\pi - \phi_2$) we have the same triangle configuration despite different chirality, and this property will bring in an additional factor of 2 in $V_{123}$. However, the chirality will not contribute twice if all the three $\vec{k}$'s are parallel to each other, i.e., $\cos \theta =\pm 1$. This explains the factor $\gamma(\cos \theta)$ we introduced in the text, which has the explicit form (also refer Ref.~\cite{ChanBlot2017} for a different derivation)
\begin{align}
    \gamma(x) = \begin{cases}
    1 & \abs{x}<1;\\
    1/2 & x = \pm 1;\\
    0 & \mathrm{else}.
    \end{cases}
\end{align}

\section{Phase-shift constraints on effective neutrino species}
\label{app:phase-shift}

\begin{table}
    \centering
    \begin{subtable}{\linewidth}
        \begin{tabularx}{\textwidth}{XXXl}
            \toprule
            Survey & P & B & P+B \\
            \midrule
         BOSS & 11 & 7.6 & $>$6.0 \\
DESI & 3.0 & 2.0 & $>$1.6 \\
Euclid & 2.4 & 1.5 & $>$1.3\\
PFS & 8.8 & 5.9 & $>$4.8  \\
SPHEREx & 4.4 & 3.1 & $>$2.5  \\
Roman (H$\alpha$) & 6.8 & 4.3 & $>$3.5\\
CVL & 0.96 & 0.59 & $>$0.47 \\
            \bottomrule
        \end{tabularx}
        \caption{Phase shift only constrains from the power spectrum and the bispectrum. {Note that the P+B constraint does not have $C^{PB}$ in consideration, thus only indicates a lower bound.}}
    \end{subtable}
    
    \begin{subtable}{\linewidth}
        \begin{tabularx}{\textwidth}{XXXl}
            \toprule
            $\alpha$-prior+Survey & P  & B  & P+B \\
            \midrule
         BOSS &  3.4 & 2.6 & $>$2.3 \\
DESI &  1.3 & 1.0 & $>$0.87 \\
Euclid &  1.0 & 0.76 & $>$0.70 \\
PFS &  2.7 & 1.9 & $>$1.6 \\
SPHEREx  & 2.1 & 1.7 & $>$1.5 \\
Roman (H$\alpha$)  & 2.0 & 1.4 & $>$1.3 \\
CVL & 0.39 & 0.27 & $>$0.23 \\
            \bottomrule
        \end{tabularx}
        \caption{Same as above, but with an $\alpha$-prior from Planck Fisher matrix. {Note that the P+B constraint does not have $C^{PB}$ in consideration, thus only indicates a lower bound.}}
    \end{subtable}

    \caption{Phase-shift only constraints on $N_{\rm eff}$ for various LSS surveys, with or without imposing a CMB prior on $\alpha$.
    }
    \label{tab:constraints-phase}
\end{table}

\begin{figure}[!htbp]
    \centering
    \includegraphics[width=\linewidth]{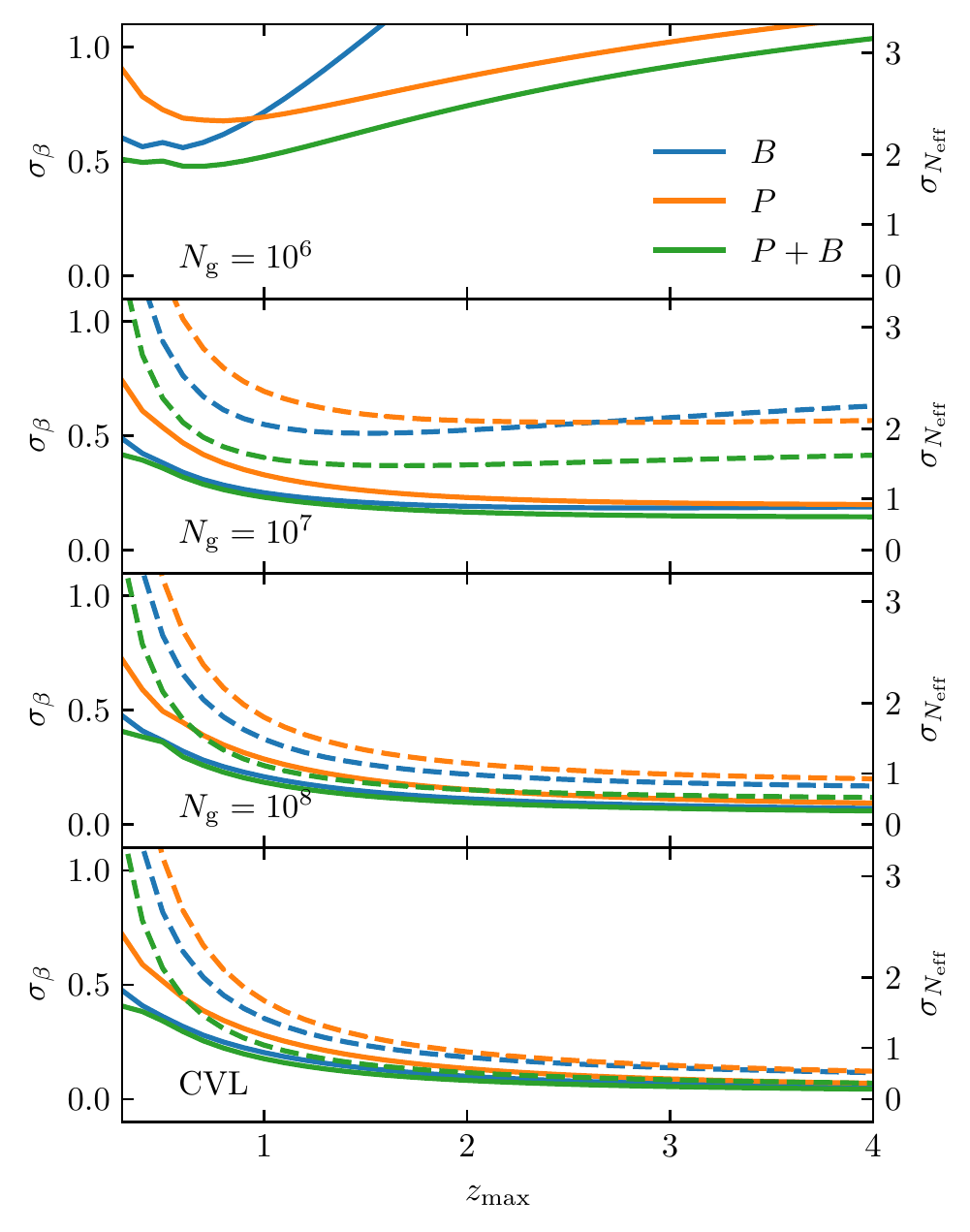}
    \caption{The 1D marginalized constraints on $\beta$ from phase-shifts only for surveys with various $z_{\rm max}$ and total galaxy density $N_g$ (see survey definitions in the text) with (solid) and without (dashed) an $\alpha$ prior from Planck 2018. We reproduce the power spectrum alone results (orange) of Ref.~\cite{BaumannGreen2018}, and additionally show the bispectrum alone (blue) and bispectrum + power spectrum (green) results.}
    \label{fig:bis-ps}
\end{figure}

In this appendix we consider a third case for the analysis where the information is solely extracted from the phase shift in the BAO wiggles. To do so, we make use of the phase shift template described in Sec. \ref{sec:background}. 

Here we only include $\alpha$ and $\beta$ as parameters and infer constraints on $N_{\rm eff}$ through $\beta(N_{\rm eff})$, to facilitate comparison with literature results (e.g. in Ref.~\cite{BaumannGreen2018}). Since $\alpha$ is redshift-dependent, there are $n_z$ values at different redshift bins, while $\beta$ is a constant over all bins. We evaluate the derivatives of $\alpha$ and $\beta$ from the analytical model shown in Eq. \eqref{equ:phase_shift}. With the reduced wavenumber $k_t$ defined as that inside $O$, we have 
\begin{align}
    \frac{\partial P_{\rm m}^{\phi}}{\partial p} = P_{\rm m }^{\rm nw}  \left.\frac{\partial O}{\partial k} \right\vert_{k=k_t} \frac{\partial k_t}{\partial p}.
\end{align}
The expression can also be extended to the galaxy power spectrum by adding additional terms referring RSD, damping, etc. However, in this prediction, we ignore multiple additional parameters like bias and polynomial terms, and keep only $\alpha$ and $\beta$.

Note that we assumed a Gaussian likelihood function for $\beta$ here, which means that the inferred $N_{\rm eff}$ distribution using the relation between $\beta$ and $N_{\rm eff}$ will be non-Gaussian (Eq.~\ref{equ:epsilon_nu}).
As a result, we define the $1\sigma$ constraint on $N_{\rm eff}$ as $\sigma_{N_{\rm eff}} \equiv N_{\rm eff}(\beta^{\rm fid}+ \sigma_\beta)-N_{\rm eff}(\beta^{\rm fid})$. 

Similarly, for the galaxy bispectrum, we obtain $\partial B^{\phi}_{\rm g}/\partial p$ through the tree level expression. %

A prior on $\alpha$ from a CMB experiment, here Planck 2018, will also be included to increase constraints on $\beta$. We choose the same definition as that of Ref.~\cite{BaumannGreen2018}. The prior matrix $C_\alpha$ is derived from  $C_\alpha^{-1} =A^{\rm T} F A$, where $F$ is the Fisher matrix of $\Lambda$CDM from Planck 2018, and $A$ is the Moore–Penrose inverse of the matrix $\nabla_{\vec \theta} \vec{\alpha}$ which is obtained with \texttt{CAMB}. %

In Fig.~\ref{fig:dist-beta} we show the $\beta$ for different surveys. Unlike in the main text, we do not include Planck constraints here. For some of the surveys like BOSS and PFS, their constraints alone without $\alpha$ prior (dashed lines) are not stringent enough to distinguish between $N_{\rm eff} = 0$ and $\infty$; but with an $\alpha$-prior from Planck 2018 (solid lines), the constraints improve considerably. These results are in agreement with Ref.~\cite{BaumannGreen2018}, where we add in addition the bispectrum results. In all the surveys, with or without the Planck prior, the bispectrum yields better constraints than the power spectrum alone, typically about 30\% improvement, with negligible further improvement when using P+B. More precise results are shown in Tab.~\ref{tab:constraints-phase} tabulates the 1-$\sigma$ constraints on $N_{\rm eff}$.

In Fig.~\ref{fig:bis-ps}, we also reproduce the $\sigma_{\beta}$ resultsfor toy surveys with varying $z_{\max}$ from Ref.~\cite{BaumannGreen2018}, but showing in addition the bispectrum contraints. Using the same setup as in Ref.~\cite{BaumannGreen2018}, we use $f_{\rm sky}=0.5$ and a redshift range going from a fixed redshift lower limit $z_{\min} = 0.1$ to an upper limit $z_{\max}$, with a bin width of $\Delta z = 0.1$. The galaxies are uniformly distributed inside the comoving volume enclosed between $z_{\min}$ and $z_{\max}$, while the total number of galaxies $N_{\rm g}$ remains fixed to $10^6$, $10^7$, $10^8$, and $\infty$ for the CVL case -- so the galaxy number density is constant with redshift at a given $z_{\rm max}$, and is lower for a higher $z_{\max}$ at a fixed total $N_{\rm g}$. %
Again the bispectrum performs better than the power spectrum, except for the low galaxy number density setups (low $N_g$ and high $z_{\rm max}$) where the shot-noise dominates. We show both the results with and without $\alpha$-prior in solid and dashed lines respectively.

\end{appendix}

\bibliography{lss,non-ads}

\end{document}